\documentclass[useAMS,usenatbib,usegraphicx]{mn2e}

\title[Three Planets in 47 UMa]{A Bayesian Periodogram Finds Evidence for Three Planets in 47 Ursae Majoris}
\author[P. C. Gregory and D. A. Fischer]{Philip C. Gregory$^{1}$\thanks{E-mail:
gregory@phas.ubc.ca} and Debra A. Fischer$^{2}$\ $^{3}$\thanks{E-mail: debra.fischer@yale.edu}\\
$^{1}$Physics and Astronomy Department, University of British Columbia, 6224 Agricultural Rd., Vancouver, BC V6T 1Z1, Canada\\
$^{2}$Department of Physics and Astronomy, San Francisco State University, San Francisco, CA 9413\\
$^{3}$Department of Astronomy, Yale University, New Haven, CT 06520}

\begin{document}

\date{Mon. Not. R. Astron. Soc. 403, 731, 2010, Accepted 20 Dec. 2009, submitted 27 Aug. 2009\\The definitive version is available at www.blackwell-synergy.com}

\pagerange{\pageref{firstpage}--\pageref{lastpage}} \pubyear{2010}

\maketitle

\label{firstpage}

\begin{abstract}
A Bayesian analysis of 47 Ursae Majoris (47 UMa) radial velocity data confirms and refines the properties of two previously reported planets with periods of 1079 and 2325 days. The analysis also provides orbital constraints on an additional long period planet with a period $\sim 10000$ days. The three planet model is found to be $~ 10^{5}$ times more probable than the next most probable model which is a two planet model.
The nonlinear model fitting is accomplished with a new hybrid Markov chain Monte Carlo (HMCMC) algorithm which incorporates parallel tempering, simulated annealing and genetic crossover operations. Each of these features facilitate the detection of a global minimum in $\chi^2$. By combining all three, the HMCMC greatly increases the probability of realizing this goal. When applied to the Kepler problem it acts as a powerful multi-planet Kepler periodogram. 

The measured periods are $1078\pm2$, $2391_{-87}^{+100}$, and $14002_{-5095}^{+4018}$d, and the corresponding eccentricities are $0.032\pm 0.014$, $0.098_{-.096}^{+.047}$, and $0.16_{-.16}^{+.09}$. The results favor low eccentricity orbits for all three. Assuming the three signals (each one consistent with a Keplerian orbit) are caused by planets, the corresponding limits on planetary mass ($M \sin i$) and semi-major axis are \\ 
($2.53_{-.06}^{+.07} M_J$, $2.10\pm 0.02\rm{au}$), ($0.54\pm 0.07 M_J$, $3.6\pm 0.1\rm{au}$), and  ($1.6_{-0.5}^{+0.3} M_J$, $11.6_{-2.9}^{+2.1}\rm{au}$), respectively. Based on a three planet model, the remaining unaccounted for noise (stellar jitter) is $5.7$m s$^{-1}$. 

The velocities of model fit residuals were randomized in multiple trials and processed using a one planet version of the HMCMC Kepler periodogram. In this situation periodogram peaks are purely the result of the effective noise. The orbits corresponding to these noise induced periodogram peaks exhibit a well defined strong statistical bias towards high eccentricity. We have characterized this eccentricity bias and designed a correction filter that can be used as an alternate prior for eccentricity, to enhance the detection of planetary orbits of low or moderate eccentricity.  
 
\end{abstract}


\section{Introduction}
\label{sec:introduction}

Improvements in precision radial velocity (RV) measurements and continued monitoring are permitting the detection of lower amplitude planetary signatures. One example of the fruits of this work is the detection of a super earth in the habitable zone surrounding Gliese 581 \citep{Udry2007}. This and other remarkable successes on the part of the observers is motivating a significant effort to improve the statistical tools for analyzing radial velocity data (e.g., \citealt{LoredoChernoff2003}, \citealt{Loredo2004}, \citealt{Cumming2004}, Gregory 2005a \& b, Ford 2005 \& 2006, \citealt{FordGregory2006}, \citealt{CummingDragomir2010}). Much of the recent work has highlighted a Bayesian MCMC approach as a way to better understand parameter uncertainties and degeneracies and to compute model probabilities.

Gregory (2005a, b, c and 2007a, b, c) presented a Bayesian MCMC algorithm that makes use of parallel tempering (PT) to efficiently explore a large model parameter space starting from a random location. It is able to identify any significant periodic signal component in the data that satisfies Kepler's laws and thus functions as a Kepler periodogram~\footnote{Following on from the pioneering work on Bayesian periodograms by \citet{Jaynes1987} and \citet{Brett1988}}. This eliminates the need for a separate periodogram search for trial orbital periods which typically assume a sinusoidal model for the signal that is only correct for a circular orbit. In addition, the Bayesian MCMC algorithm provides full marginal parameters distributions for all the orbital elements that can be determined from radial velocity data. The algorithm includes an innovative two stage adaptive control system that automates the selection of efficient Gaussian parameter proposal distributions.

The latest version of the algorithm, \citet{Gregory2009}, incorporates a genetic crossover operation into the MCMC algorithm. The new adaptive hybrid MCMC algorithm (HMCMC) incorporates parallel tempering, simulated annealing and genetic crossover operations. Each of these techniques was designed to facilitate the detection of a global minimum in $\chi^2$. Combining all three in an adaptive hybrid MCMC greatly increase the probability of realizing this goal. 

\citet{ButlerMarcy1996} first reported a 1090 day companion to 47 UMa using data from Lick Observatory. With additional velocity measurements over 13 yr, \citet{Fischer2002} announced
a long-period second planet, 47 UMa c, with a period of $2594\pm90$ days and a mass of $0.76 M_J$. \citet{Naef2004} reported observations from the fiber fed echelle spectrograph ELODIE of 47 UMa, and noted that the second
planet was not evident in their data.  
\citet{Wittenmyer2007} reported that there is still substantial ambiguity as to the orbital parameters of the proposed planetary companion 47 UMa c. They gave a period of 7586 day for one orbital solution, and 2594 day for two others. In their latest work \citet{Wittenmyer2009}, their
best-fit 2-planet model now calls for $P_2 = 9660$ days. In this paper we present a Bayesian analysis of the latest Lick observatory measurements and a combined Lick plus McDonald Observatory (\citealt{Wittenmyer2009}) data set.

We also report on an investigation of the behavior of the Bayesian HMCMC Kepler periodogram to noise. The noise data sets were formed by randomly interchanging velocity measurements.

\section{The adaptive hybrid MCMC}

The adaptive hybrid MCMC (HMCMC) is a very general Bayesian nonlinear model fitting program. After specifying the nonlinear model, data and priors, Bayes theorem dictates the target joint probability distribution for the model parameters which can be very complex. To compute the marginals for any subset of the parameters it is necessary to integrate the joint probability distribution over the remaining parameters. In high dimensions, the principle tool for carrying out the integrals is Markov chain Monte Carlo based on the Metropolis algorithm. The greater efficiency of an MCMC stems from its ability, after an initial burn-in period, to generate  samples in parameter space in direct proportion to the joint target probability distribution. In contrast, straight Monte Carlo integration randomly samples the parameter space and wastes most of its time sampling regions of very low probability. 

An important feature that prevents the HMCMC from becoming stuck in a local probability maximum is parallel tempering. Multiple MCMC chains are run in parallel. The joint probability density distribution for the parameters ($\vec{X}$) of model $M_i$, for a particular chain, is given by
\begin{equation}
p(\vec{X}|D,M_i,I,\beta) \propto p(\vec{X}|M_i,I)\times p(D|\vec{X},M_i,I)^{\beta}.
\label{eq:tempering}
\end{equation}
Each MCMC chain corresponding to a different $\beta$, with the value of $\beta$ ranging from zero to 1. When the exponent $\beta = 1$, the term on the LHS of the equation is the target joint probability distribution for the model parameters, $p(\vec{X}|D,M_i,I)$. It is the posterior probability of a particular choice of parameter vector, $\vec{X}$, given the data represented by $D$, the model choice $M_i$, and the prior information $I$. In general, the model parameter space of interest is a continuum so $p(\vec{X}|D,M_i,I)$ is a probability density distribution. The first term on the RHS of the equation, $p(\vec{X}|M_i,I)$, is the prior probability density distribution of $\vec{X}$, prior to the consideration of the current data $D$. The second term, $p(D|\vec{X}M_i,I)$, is called the likelihood and it is the probability that we would have obtained the measured data $D$ for this particular choice of parameter vector $\vec{X}$, model $M_i$, and prior information $I$. At the very least, the prior information, $I$, must specify the class of alternative models (hypotheses) being considered (hypothesis space of interest) and the relationship between the models and the data (how to compute the likelihood). For further details of the likelihood function for this problem see \cite{Gregory2005b}. In many situations the log of the likelihood is simply proportional to the familiar $\chi^2$ statistic. If we later acquire another data set $D^{\prime}$ then the new prior, $p(\vec{X}|M_i,I^{\prime})$, is equal to our previous posterior, $p(\vec{X}|D,M_i,I)$, i.e., $I^{\prime} = I,D$. 
An exponent $\beta = 0$, yields a broader joint probability density equal to the prior. The reciprocal of $\beta$ is analogous to a temperature, the higher the temperature the broader the distribution. 

For parameter estimation purposes 8 chains\\
($\beta=\{0.09, 0.13, 0.20, 0.29, 0.39, 0.52, 0.72, 1.0\}$) were employed. At an intervals of 10 iterations, a pair of adjacent chains on the tempering ladder are chosen at random and a proposal made to swap their parameter states. A Monte Carlo acceptance rule determines the probability for the proposed swap to occur (e.g., Gregory 2005a, eq. 12.12). This swap allows for an exchange of information across the population of parallel simulations. In low $\beta$ (higher temperature) simulations, radically different configurations can arise, whereas in higher $\beta$ (lower temperature) states, a configuration is given the chance to refine itself. The lower $\beta$ chains can be likened to a series of scouts that explore the parameter terrain on different scales. The final samples are drawn from the $\beta  = 1$ chain, which corresponds to the desired target probability distribution. For $\beta \ll 1$, the distribution is much flatter. The choice of $\beta$ values can be checked by computing the swap acceptance rate. When they are too far apart the swap rate drops to very low values.

Each parallel chain employs the Metropolis algorithm. At each iteration a proposal to jump to a new location in parameter space is generated from independent Gaussian proposal distributions (centered on the current parameter location), one for each parameter. In general, the $\sigma$'s of these Gaussian proposal distributions are different because the parameters can be very different entities. Also if the $\sigma$'s are chosen too small, successive samples will be highly correlated and will require many iterations to obtain an equilibrium set of samples. If the $\sigma$'s are too large, then proposed samples will very rarely be accepted. The process of choosing a set of useful proposal $\sigma$'s when dealing with a large number of different parameters can be very time consuming. In parallel tempering MCMC, this problem is compounded because of the need for a separate set of Gaussian proposal $\sigma$'s for each chain (different tempering levels). This process is automated by an innovative two stage statistical control system (\citealt{Gregory2007b}, \citealt{Gregory2009}) in which the error signal is proportional to the difference between the current joint parameter acceptance rate and a target acceptance rate, typically 25\% (\citealt{Roberts1997}). A schematic of the full adaptive control system (CS) is shown in 
Figure 1.
\begin{figure*}
\begin{center}
\includegraphics[width=140mm]{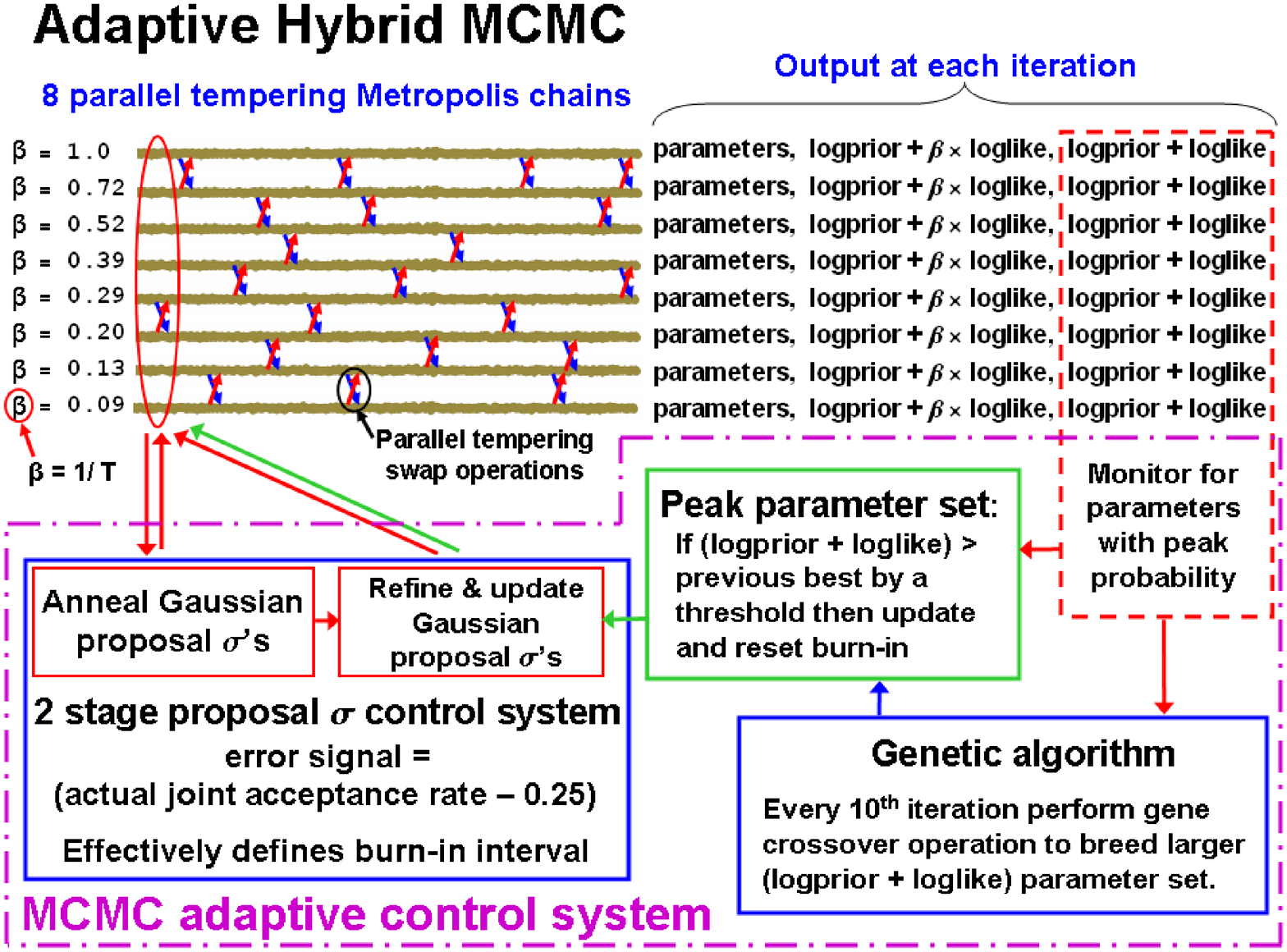}
\caption{Schematic of the operation of the adaptive hybrid MCMC algorithm.}
\end{center}
\label{fig:schem}
\end{figure*}
 
The first stage CS, which involves annealing the set of Gaussian proposal distribution $\sigma$'s, was described in Gregory 2005a. An initial set of proposal $\sigma$'s ($\approx 10\%$ of the prior range for each parameter) are used for each chain. During the major cycles, the joint acceptance rate is measured based on the current proposal $\sigma$'s and compared to a target acceptance rate. During the minor cycles, each proposal $\sigma$ is separately perturbed to determine an approximate gradient in the acceptance rate for that parameter. The $\sigma$'s are then jointly modified by a small increment in the direction of this gradient. This is done for each of the parallel chains. Proposals to swap parameter values between chains are allowed during major cycles but not within minor cycles. 

\begin{figure}
\includegraphics[width=85mm]{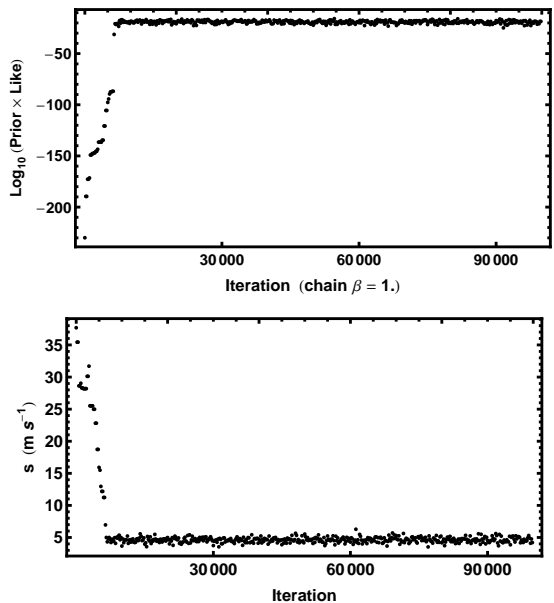}
\caption{The upper panel is a plot of the Log$_{10}$[Prior $\times$ Likelihood] versus MCMC iteration. The lower panel is a similar plot for the extra noise term $s$. Initially $s$ is inflated and then rapidly decays to a much lower level as the best fit parameter values are approached.}
\label{fig:sAnnealing}
\end{figure} 

The annealing of the proposal $\sigma$'s occurs while the MCMC is homing in on any significant peaks in the target probability distribution. Concurrent with this, another aspect of the annealing operation takes place whenever the Markov chain is started from a location in parameter space that is far from the best fit values. This automatically arises because all the models considered incorporate an extra additive noise \citet{Gregory2005b}, for reasons discussed in Section~\ref{sec:analysis}, whose probability distribution is Gaussian with zero mean and with an unknown standard deviation $s$. When the $\chi^2$ of the fit is very large, the Bayesian Markov chain automatically inflates $s$ to include anything in the 
data that cannot be accounted for by the model with the current set of 
parameters and the known measurement errors. This results in a smoothing out of the detailed structure in the $\chi^2$ surface and, as pointed out by \citet{Ford2006}, allows the Markov chain to explore the large scale structure in parameter space more quickly. The chain begins to decrease the value of $s$ as it settles in near the best-fit parameters. An example of this is shown in 
Figure~\ref{fig:sAnnealing}. In the early stages $s$ is inflated to around 38 m s$^{-1}$ and then decays to a value of $\approx 4$ m s$^{-1}$ over the first 9,000 iterations. This is similar to simulated annealing, but does not require choosing a cooling scheme. 

Although the first stage CS achieves the desired joint acceptance rate, it often happens that a subset of the proposal $\sigma$'s are too small leading to an excessive autocorrelation in the MCMC iterations for these parameters. Part of the second stage CS corrects for this. 
The goal of the second stage is to achieve a set of proposal $\sigma$'s that equalizes the MCMC acceptance rates when new parameter values are proposed separately and achieves the desired acceptance rate when they are proposed jointly. Details of the second stage CS were given in \citealt{Gregory2007b}.

The first stage is run only once at the beginning, but the second stage can be executed repeatedly, whenever a significantly improved parameter solution emerges. 
Frequently, the burn-in period occurs within the span of the first stage CS, i.e., the significant peaks in the joint parameter probability distribution are found, and the second stage improves the choice of proposal $\sigma$'s based on the highest probability parameter set. 
Occasionally, a new higher (by a user specified threshold) target probability parameter set emerges after the first two stages of the CS are completed. The control system has the ability to detect this and automatically re-activate the second stage. In this sense the CS is adaptive. If this happens the iteration corresponding to the end of the control system is reset.
The useful MCMC simulation data is obtained after the CS are switched off.

The adaptive capability of the control system can be appreciated from an examination of Figure 1. The upper left portion of the figure depicts the MCMC iterations from the 8 parallel chains, each corresponding to a different tempering level $\beta$ as indicated on the extreme left. One of the outputs obtained from each chain at every iteration (shown at the far right) is the $\log$ prior $+ \log$ likelihood. This information is continuously fed to the CS which constantly updates the most probable parameter combination regardless of which chain the parameter set occurred in. This is passed to the "Peak parameter set" block of the CS. Its job is to decide if a significantly more probable parameter set has emerged since the last execution of the second stage CS. If so, the second stage CS is re-run using the new more probable parameter set which is the basic adaptive feature of the CS.

The CS also includes genetic algorithm block which is shown in the bottom right of Figure 1. The current parameter set can be treated as a set of genes. In the present version, one gene consists of the parameter set that specify one orbit. On this basis, a three planet model has three genes. At any iteration there exist within the CS the most probable parameter set to date $\vec{X}_{\rm max}$, and the most probable parameter set from the 8 chains for the most recent iteration $\vec{X}_{\rm cur}$. At regular intervals (user specified) each gene from $\vec{X}_{\rm cur}$ is swapped for the corresponding gene in $\vec{X}_{\rm max}$. If either substitution leads to a higher probability it is retained and $\vec{X}_{\rm max}$ updated. The effectiveness of this operation can be tested by comparing the number of times the gene crossover operation gives rise to a new value of $\vec{X}_{\rm max}$ compared to the number of new $\vec{X}_{\rm max}$ arising from the normal parallel tempering MCMC iterations. The gene crossover operations prove to be very effective, and give rise to new $\vec{X}_{\rm max}$ values $\approx 3$ times more often than MCMC operations. Of course, most of these swaps lead to very minor changes in probability but occasionally big jumps are created. 

Gene swaps from $\vec{X}_{\rm cur2}$, the parameters of the second most probable current chain, to $\vec{X}_{\rm max}$ are also utilized. This gives rise to new values of $\vec{X}_{\rm max}$ at a rate approximately half that of swaps from $\vec{X}_{\rm cur}$ to $\vec{X}_{\rm max}$. Crossover operations at a random point in the entire parameter set did not prove as effective except in the single planet case where there is only one gene. Further experimentation with this concept is ongoing.    

\section{Data and Analysis}
\label{sec:analysis}

Our initial analysis was based on data obtained at the Lick observatory and spans a period of 21.6 years. The data are listed in Tables~\ref{tab:data1} and \ref{tab:data2}. In addition to the observation time, radial velocity $(RV)$, and velocity error $(\Delta RV)$, the detector dewar number used is also included. We originally analyzed the data ignoring possible residual velocity offsets associated with dewar changes (Case A). To investigate how robust the results were we subsequently repeated the analysis incorporating the dewar velocity offsets as additional unknown parameters (Case B). In Case A the data from all 6 dewars are used. For Case B we excluded dewar 1 because with only a single measurement the analysis is unable to separate the offset from the model velocity contribution which reduces the time base by 235 days. Results for the two cases follow in subsequent sections labeled accordingly. In Section~\ref{sec:discussion}, we extend the analysis to include the \citet{Wittenmyer2009} data from the 9.2 m Hobby-Eberly Telescope (HET) and 2.7 m Harlam J. Smith (HJS) telescopes of the McDonald Observatory. In the rest of this section we describe the model fitting equations and the selection of priors for the model parameters. We also characterize a noise induced eccentricity bias that leads to a second choice for an eccentricity prior.

\begin{table*}
 \centering
 \begin{minipage}{150mm}
  \caption{Radial velocities (RV) for 47 UMa. The $\Delta$RV column gives the RV uncertainty and the next column gives the detector dewar number.}
  \label{tab:data1}
\small
  \begin{tabular}{@{}lllclllclllc@{}}
  \hline
   JD-2440000 & RV & $\Delta$RV & Dewar & JD-2440000 & RV & $\Delta$RV & Dewar & JD-2440000 & RV & $\Delta$RV & Dewar\\
& m s$^{-1}$ & m s$^{-1}$ & & & m s$^{-1}$ & m s$^{-1}$ & & &m s$^{-1}$ & m s$^{-1}$ & \\ 
 \hline
  6959.7372 &  -40.70 &   14.00 & 1 & 11607.9163 &  -17.77 &    4.51 & 18 & 12722.8295 &  -20.88 &    3.13 & 24\\
  7194.9122 &  -33.96 &    7.49 & 6 & 11626.7707 &  -34.76 &    6.65 & 18 & 12737.7703 &  -10.01 &    2.47 & 24\\
  7223.7982 &  -18.31 &    6.14 & 6 & 11627.7539 &  -29.07 &    5.87 & 18 & 12793.7298 &    1.53 &    2.41 & 24\\
  7964.8927 &   20.40 &    8.19 & 6 & 11628.7275 &  -34.86 &    5.71 & 18 & 12794.7134 &   -5.06 &    2.20 & 24\\
  8017.7302 &   -8.18 &   10.57 & 6 & 11629.8320 &  -32.06 &    4.48 & 18 & 12834.6981 &   21.08 &    2.83 & 24\\
  8374.7707 &  -20.25 &    9.37 & 6 & 11700.6937 &   -2.83 &    4.80 & 18 & 12991.0537 &   57.90 &    3.94 & 24\\
  8647.8971 &   62.95 &   11.41 & 8 & 11861.0498 &   36.20 &    5.53 & 18 & 12992.0732 &   55.57 &    4.69 & 24\\
  8648.9100 &   51.93 &   11.02 & 8 & 11874.0684 &   39.39 &    5.34 & 18 & 13009.0525 &   53.57 &    2.70 & 24\\
  8670.8777 &   74.56 &   11.45 & 8 & 11881.0443 &   32.79 &    4.41 & 18 & 13009.9546 &   51.65 &    2.88 & 24\\
  8745.6907 &   71.89 &    8.76 & 8 & 11895.0663 &   33.89 &    4.28 & 18 & 13018.9971 &   55.32 &    4.48 & 24\\
  8992.0612 &   23.42 &   11.21 & 8 & 11906.0148 &   34.69 &    3.91 & 18 & 13020.9531 &   39.96 &    5.42 & 24\\
  9067.7708 &    4.86 &    7.00 & 8 & 11907.0112 &   37.74 &    4.24 & 18 & 13022.0027 &   46.17 &    5.15 & 24\\
  9096.7339 &   -6.19 &    6.79 & 8 & 11909.0420 &   39.07 &    3.76 & 18 & 13044.9198 &   58.89 &    3.33 & 24\\
  9122.6909 &  -27.90 &    7.91 & 8 & 11910.9537 &   36.96 &    4.13 & 18 & 13068.8447 &   54.81 &    5.38 & 24\\
  9172.6855 &  -18.68 &   10.55 & 8 & 11914.0674 &   34.35 &    5.17 & 18 & 13069.8323 &   48.36 &    3.34 & 24\\
  9349.9122 &  -32.93 &    9.52 & 8 & 11915.0473 &   41.14 &    3.72 & 18 & 13072.8875 &   45.63 &    2.93 & 24\\
  9374.9638 &  -29.14 &    8.67 & 8 & 11916.0335 &   40.99 &    3.47 & 18 & 13078.8069 &   52.75 &    3.30 & 24\\
  9411.8387 &  -16.88 &   12.81 & 8 & 11939.9703 &   42.47 &    4.72 & 18 & 13079.8275 &   52.69 &    3.18 & 24\\
  9481.7197 &  -33.01 &   13.40 & 8 & 11946.9598 &   42.21 &    4.19 & 18 & 13080.7919 &   52.88 &    3.27 & 24\\
  9767.9184 &   64.68 &    5.34 & 39 & 11969.9024 &   48.36 &    4.29 & 18 & 13081.8171 &   48.72 &    2.99 & 24\\
  9768.9072 &   62.32 &    4.79 & 39 & 11971.8934 &   52.56 &    4.80 & 18 & 13100.8148 &   53.34 &    3.83 & 24\\
  9802.7911 &   63.99 &    3.61 & 39 & 11998.7785 &   49.07 &    3.81 & 18 & 13107.7773 &   35.01 &    4.43 & 24\\
 10058.0797 &   32.21 &    3.18 & 39 & 11999.8203 &   48.13 &    3.98 & 18 & 13119.7426 &   50.57 &    4.24 & 24\\
 10068.9773 &   36.13 &    4.01 & 39 & 12000.8587 &   50.97 &    4.16 & 18 & 13120.6914 &   42.03 &    2.83 & 24\\
 10072.0117 &   38.76 &    4.10 & 39 & 12028.7386 &   60.65 &    4.39 & 18 & 13131.6826 &   51.29 &    4.13 & 24\\
 10088.9932 &   23.38 &    3.54 & 39 & 12033.7461 &   49.37 &    4.93 & 18 & 13132.7334 &   38.98 &    5.04 & 24\\
 10089.9473 &   26.18 &    3.19 & 39 & 12040.7593 &   47.52 &    3.54 & 18 & 13147.6943 &   44.95 &    4.94 & 24\\
 10091.9004 &   18.37 &    4.23 & 39 & 12041.7192 &   49.30 &    3.37 & 18 & 13155.7006 &   38.98 &    2.92 & 24\\
 10120.9179 &   17.53 &    3.91 & 39 & 12042.6957 &   45.95 &    3.88 & 18 & 13156.7062 &   40.23 &    2.77 & 24\\
 10124.9042 &   23.41 &    3.69 & 39 & 12071.7291 &   53.86 &    9.39 & 18 & 13157.6869 &   43.30 &    2.98 & 24\\
 10125.8234 &   18.49 &    3.61 & 39 & 12073.7217 &   44.45 &    4.61 & 18 & 13339.0682 &    9.72 &    3.31 & 24\\
 10127.8979 &   13.98 &    3.77 & 39 & 12101.6865 &   59.74 &    6.62 & 18 & 13363.0139 &    2.22 &    4.52 & 24\\
 10144.8770 &   13.75 &    4.67 & 39 & 12103.6875 &   41.81 &    5.71 & 18 & 13363.9655 &   -9.51 &    4.21 & 24\\
 10150.7964 &   12.07 &    3.89 & 39 & 12104.6855 &   47.90 &    5.78 & 18 & 13383.9778 &  -11.43 &    9.04 & 24\\
 10172.8289 &    4.69 &    4.13 & 39 & 12105.6836 &   41.69 &    5.71 & 18 & 13385.0057 &  -23.57 &    4.17 & 24\\
 10173.7627 &    9.36 &    5.29 & 39 & 12216.0355 &   27.62 &    4.56 & 18 & 13385.9946 &  -25.20 &    3.94 & 24\\
 10181.7425 &   -2.47 &    3.18 & 39 & 12222.0432 &   28.69 &    4.35 & 18 & 13388.0012 &  -19.02 &   10.53 & 24\\
 10187.7390 &    7.94 &    4.22 & 39 & 12278.0718 &   -6.78 &    4.79 & 18 & 13389.9276 &  -32.39 &    4.48 & 24\\
 10199.7291 &    5.49 &    3.62 & 39 & 12279.0680 &   -2.81 &    4.54 & 18 & 13390.9468 &  -18.25 &    4.75 & 24\\
 10203.7330 &    1.63 &    4.23 & 39 & 12283.0395 &    2.35 &    7.53 & 18 & 13391.9987 &  -30.29 &    4.56 & 24\\
 10214.7308 &   -2.09 &    3.54 & 39 & 12286.0614 &   -4.09 &    3.53 & 18 & 13392.9238 &  -31.99 &    4.95 & 24\\
 10422.0176 &  -32.32 &    4.05 & 39 & 12288.0176 &   -3.07 &    4.86 & 18 & 13402.9585 &  -13.79 &    4.74 & 24\\
 10438.0010 &  -23.92 &    4.30 & 39 & 12306.9303 &  -20.54 &    6.38 & 18 & 13403.9527 &  -23.51 &    4.68 & 24\\
 10442.0273 &  -26.34 &    3.84 & 39 & 12314.9275 &  -15.06 &    3.57 & 18 & 13404.9472 &  -24.04 &    5.42 & 24\\
 10502.8535 &  -15.99 &    3.86 & 39 & 12315.9273 &  -12.71 &    2.72 & 18 & 13436.7878 &  -24.91 &    5.44 & 24\\
 10504.8594 &  -19.78 &    4.24 & 39 & 12316.9996 &   -0.12 &    6.13 & 18 & 13437.8865 &  -40.32 &    5.46 & 24\\
 10536.8441 &   -3.96 &    4.58 & 39 & 12348.8617 &  -22.34 &    4.03 & 18 & 13438.8413 &  -31.99 &    3.91 & 24\\
 10537.8426 &   -6.81 &    3.81 & 39 & 12375.7996 &  -26.80 &    3.35 & 24 & 13439.8543 &  -36.26 &    4.13 & 24\\
 10563.6734 &   -0.73 &    3.76 & 39 & 12376.7234 &  -28.65 &    3.71 & 24 & 13440.7724 &  -32.85 &    5.39 & 24\\
 10579.6952 &   11.11 &    3.55 & 39 & 12380.7568 &  -28.60 &    3.90 & 18 & 13441.8656 &  -34.10 &    5.38 & 24\\
 10610.7188 &   12.05 &    3.34 & 39 & 12388.7530 &  -34.84 &    3.12 & 24 & 13460.8047 &  -39.69 &    4.47 & 24\\
 10793.9570 &   58.79 &    3.97 & 39 & 12389.7036 &  -43.22 &    3.76 & 24 & 13475.7043 &  -39.74 &    4.67 & 24\\
 10795.0391 &   62.55 &    4.07 & 39 & 12577.0504 &  -37.38 &    3.58 & 24 & 13476.7068 &  -39.51 &    4.68 & 24\\
 10978.6848 &   55.48 &    4.51 & 18 & 12599.0475 &  -33.96 &    2.53 & 24 & 13477.7253 &  -38.11 &    4.41 & 24\\
 11131.0654 &   37.48 &    6.35 & 18 & 12609.0665 &  -38.85 &    3.56 & 24 & 13478.7598 &  -43.63 &    4.14 & 24\\
 11175.0273 &   21.32 &    7.24 & 18 & 12631.9926 &  -26.01 &    3.61 & 24 & 13479.7748 &  -47.47 &    4.21 & 24\\
 11242.8418 &    1.34 &    4.82 & 18 & 12657.0184 &  -41.11 &    2.29 & 24 & 13511.7132 &  -53.62 &    4.11 & 24\\
 11303.7119 &  -25.20 &    4.20 & 18 & 12687.8597 &  -26.21 &    3.48 & 24 & 13512.6881 &  -40.62 &    4.35 & 24\\
 11508.0703 &  -36.52 &    8.34 & 18 & 12688.9015 &  -34.23 &    5.04 & 24 & 13744.0283 &  -38.93 &    4.31 & 24\\
 11536.0640 &  -43.83 &    4.75 & 18 & 12705.8382 &  -25.21 &    2.74 & 24 & 13744.9815 &  -40.34 &    4.30 & 24\\
\hline
\end{tabular}
\normalsize
\end{minipage}
\end{table*} 
\begin{table*}
 \centering
 \begin{minipage}{150mm}
  \caption{Radial velocities (RV) for 47 UMa. The $\Delta$RV column gives the RV uncertainty and the next column gives the detector dewar number.}
  \label{tab:data2}
\small
  \begin{tabular}{@{}lllclllclllc@{}}
  \hline
   JD-2440000 & RV & $\Delta$RV & Dewar & JD-2440000 & RV & $\Delta$RV & Dewar & JD-2440000 & RV & $\Delta$RV & Dewar\\
& m s$^{-1}$ & m s$^{-1}$ & & & m s$^{-1}$ & m s$^{-1}$ & & &m s$^{-1}$ & m s$^{-1}$ & \\ 
 \hline
 13753.0361 &  -53.52 &    2.95 & 24 & 14135.8630 &   24.39 &    2.32 & 24 & 14598.7489 &  -40.52 &    3.10 & 24\\
 13755.8982 &  -41.94 &    5.04 & 24 & 14165.8471 &   30.47 &    3.34 & 24 & 14622.7505 &  -41.29 &    2.42 & 24\\
 13773.8466 &  -51.29 &    4.91 & 24 & 14196.8162 &   35.39 &    3.14 & 24 & 14623.7115 &  -34.55 &    2.52 & 24\\
 13866.7278 &  -21.49 &    4.38 & 24 & 14219.7662 &   24.68 &    3.08 & 24 & 14784.0515 &  -31.13 &    5.10 & 24\\
 13867.7226 &  -27.25 &    4.67 & 24 & 14220.7881 &   33.23 &    3.26 & 24 & 14785.0826 &  -34.05 &    4.85 & 24\\
 13868.7523 &  -25.55 &    4.56 & 24 & 14253.6937 &   27.72 &    2.73 & 24 & 14845.0201 &  -30.72 &    1.74 & 24\\
 13869.7295 &  -13.48 &    4.15 & 24 & 14254.7002 &   24.49 &    2.65 & 24 & 14847.9355 &  -26.93 &    3.42 & 24\\
 14074.0693 &   34.48 &    3.23 & 24 & 14427.0782 &   -6.25 &    4.40 & 24 & 14848.9727 &  -30.86 &    2.74 & 24\\
 14099.0854 &   40.26 &    3.15 & 24 & 14450.0617 &  -10.65 &    3.41 & 24 & 14849.9710 &  -27.88 &    3.09 & 24\\
 14100.0667 &   32.10 &    3.21 & 24 & 14462.0257 &  -16.81 &    2.42 & 24 & 14850.9698 &  -31.85 &    3.06 & 24\\
 14102.0466 &   36.94 &    3.38 & 24 & 14547.9127 &  -27.51 &    3.24 & 24 & 14863.9813 &  -27.92 &    4.26 & 24\\
 14104.0288 &   36.91 &    4.44 & 24 & 14574.8034 &  -52.51 &    1.79 & 24 & 14864.9193 &  -29.72 &    5.10 & 24\\
 14133.9656 &   32.61 &    4.66 & 24 & 14578.8416 &  -41.13 &    2.11 & 24 & 14865.9624 &  -19.55 &    5.54 & 24\\
 14134.9264 &   25.80 &    2.71 & 24 &  &  &  &  &  &  &  & \\
\hline
\end{tabular}
\normalsize
\end{minipage}
\end{table*} 

We have investigated the 47 UMa data using models ranging from a single planet to five planets. For a one planet model the predicted radial velocity is given by
\begin{equation}
v(t_i) = V + K [\cos\{\theta(t_i+\chi P)+\omega\} + e \cos \omega],
\label{eq:orbit1}
\end{equation}
and involves the 6 unknown parameters
\begin{itemize}
\item[] $V =$ a constant velocity.
\item[] $K =$ velocity semi-amplitude. 
\item[] $P =$ the orbital period.
\item[] $e =$ the orbital eccentricity.
\item[] $\omega =$ the longitude of periastron.
\item[] $\chi =$ the fraction of an orbit, prior to the start of data taking, that periastron occurred at. Thus, $\chi P =$ the number of days prior to $t_i = 0$ that the star was at periastron, for an orbital period of P days. 
\item[] $\theta(t_i+\chi P) =$ the true anomaly, the angle of the star in its orbit relative to periastron at time $t_i$.
\end{itemize}

We utilize this form of the equation because we obtain the dependence of $\theta$ on $t_i$ by solving the conservation of angular momentum equation
\begin{equation}
\frac{d\theta}{dt} - \frac{2 \pi [1+e\cos \theta(t_i+\chi \; P)]^2}{P (1-e^2)^{3/2}} = 0.
\label{eq:orbit2}
\end{equation}
Our algorithm is implemented in {\it Mathematica} and it proves faster for {\it Mathematica} to solve this differential equation than solve the equations relating the true anomaly to the mean anomaly via the eccentric anomaly. {\it Mathematica} generates an accurate interpolating function between $t$ and $\theta$ so the differential equation does not need to be solved separately for each $t_i$. Evaluating the interpolating function for each $t_i$ is very fast compared to solving the differential equation, so the algorithm should be able to handle much larger samples of radial velocity data than those currently available without a significant increase in computational time.
For example, an increase in the data by a factor of 6.5 resulted in only an 18\% increase in execution time.

As described in more detail in \citealt{Gregory2007a}, we employed a re-parameterization of $\chi$ and $\omega$ to improve the MCMC convergence speed motivated by the work of Ford (2006). The two new parameters are $\psi=2\pi\chi+\omega$ and $\phi=2 \pi\chi-\omega$. Parameter $\psi$ is well determined for all eccentricities. Although $\phi$ is not well determined for low eccentricities, it is at least orthogonal to the $\psi$ parameter. We use a uniform prior for $\psi$ in the interval 0 to $4 \pi$ and uniform prior for $\phi$ in the interval $-2 \pi$ to $+2 \pi$. This insures that a prior that is wraparound continuous in $(\chi,\omega)$ maps into a wraparound continuous distribution in $(\psi,\phi)$. The big $(\psi,\phi)$ square holds two copies of the probability patch in $(\chi,\omega)$ which doesn't matter. What matters is that the prior is now wraparound continuous in $(\psi,\phi)$. 

In a Bayesian analysis we need to specify a suitable prior for each parameter. These are tabulated in Table~\ref{tab:priors}. For the current problem, the prior given in Equation~\ref{eq:tempering} is the product of the individual parameter priors. Detailed arguments for the choice of each prior were given in \citealt{Gregory2007a}.     
\begin{table*}
 \centering
 \begin{minipage}{140mm}
  \caption{Prior parameter probability distributions.}
  \label{tab:priors}
  \begin{tabular}{@{}llll@{}}
  \hline
   Parameter    &    Prior        & Lower bound & Upper bound\\
 \hline
Orbital frequency  & $p(\ln f_1, \ln f_2, \cdots \ln f_n|M_n,I) = \frac{n!}{[\ln (f_H/f_L)]^n}$  & 1/1.5 d & 1/1000 yr  \\
&\ ($n=$number of planets)  & &  \\
& & & \\
Velocity $K_i$  &  Modified Jeffreys~\footnote{Since the prior lower limits for $K$ and $s$ include zero, we used a modified Jeffreys prior of the form
\begin{equation}
p(X|M,I) = \frac{1}{X+X_0}\; \frac{1}{\ln\left(1+\frac{X_{\rm max}}{X_0}\right)}
\label{eq:orbit13}
\end{equation}
For $X \ll X_0$, $p(X|M,I)$ behaves like a uniform prior and for $X \gg X_0$ it behaves like a Jeffreys prior. The $\ln\left(1+\frac{ X_{\rm max}}{X_0}\right)$ term in the denominator ensures that the prior is normalized in the interval 0 to $X_{\rm max}$.} & 0 \ (K$_0 = 1)$ &  $K_{\rm max}\ \left(\frac{P_{\rm min}}{P_i}\right)^{1/3} \frac{1}{\sqrt{1-e_i^2}}$ \\
\ \ \  (m s$^{-1}$) & & & \\
  & \ \ \ \ \ $\frac{(K+K_0)^{-1}}{\ln{\left[1+\frac{K_{\rm max}}{K_0} \ \left(\frac{P_{\rm min}}{P_i}\right)^{1/3} \frac{1}{\sqrt{1-e_i^2}}\right]}}$
 &  & $K_{\rm max}=2129$\\
 & & & \\
V  (m s$^{-1}$) & Uniform & $-K_{\rm max}$ & $K_{\rm max}$  \\
& & & \\
$e_i$ Eccentricity & a) Uniform & 0 & 1 \\
 & b) Ecc. noise bias correction filter& 0 & 0.99 \\
& & & \\
& & & \\
$\omega_i$ Longitude of & Uniform & $0$ & $2 \pi$ \\
\ \ \ \ periastron &  &  & \\
& & & \\
$s$ Extra noise   (m s$^{-1}$) & $\frac{(s+s_0)^{-1}}{\ln{\left(1+\frac{s_{\rm max}}{s_{0}}\right)}}$ & 0  \ (s$_0 = 1$)& $K_{\rm max}$  \\
\hline
\end{tabular}
\end{minipage}
\end{table*}

\citealt{Gregory2007a} discussed two different strategies to search the orbital frequency parameter space for a multi-planet model: (i) an upper bound on $f_1 \le f_2 \le \cdots \le f_n$  is utilized to maintain the identity of the frequencies, and (ii) all $f_i$ are allowed to roam over the entire frequency range  and the parameters re-labeled afterwards. Case (ii) was found to be significantly more successful at converging on the highest posterior probability peak in fewer iterations during repeated blind frequency searches. In addition, case (ii) more easily permits the identification of two planets in 1:1 resonant orbits. We adopted approach (ii) in the current analysis. 

All of the models considered in this paper incorporate an extra noise parameter, $s$, that can allow for any additional noise beyond the known measurement uncertainties~\negthinspace\footnote{In the absence of detailed knowledge of the sampling distribution for the extra noise, we pick a Gaussian because for any given finite noise variance it is the distribution with the largest uncertainty as measured by the entropy, i.e., the maximum entropy distribution (\citealt{Jaynes1957}, \citealt{Gregorybook} section 8.7.4.)}. We assume the noise variance is finite and adopt a Gaussian distribution with a variance $s^2$. Thus, the combination of the known errors and extra noise has a Gaussian distribution with variance $= \sigma_i^2 + s^2$, where $\sigma_i$ is the standard deviation of the known noise for i$^{\mbox{\tiny th}}$ data point. For example, suppose that the star actually has two planets, and the model assumes only one is present. In regard to the single planet model, the velocity variations induced by the unknown second planet acts like an additional unknown noise term. Other factors like star spots and chromospheric activity can also contribute to this extra velocity noise term which is often referred to as stellar jitter. Several researchers have attempted to estimate stellar jitter for individual stars based on statistical correlations with observables (e.g., \citealt{Saar1997}, \citealt{Saar1998}, \citealt{Wright2005}). In general, nature is more complicated than our model and known noise terms. Marginalizing $s$ has the desirable effect of treating anything in the data that can't be explained by the model and known measurement errors as noise, leading to conservative estimates of orbital parameters. See Sections 9.2.3 and 9.2.4 of \citet{Gregorybook} for a tutorial demonstration of this point. If there is no extra noise then the posterior probability distribution for $s$ will peak at $s = 0$. The upper limit on $s$ was set equal to $K_{\rm max}$. We employed a modified Jeffrey's prior for $s$ with a knee, $s_0 = 1$m s$^{-1}$. 
\begin{figure}
\includegraphics[width=80mm]{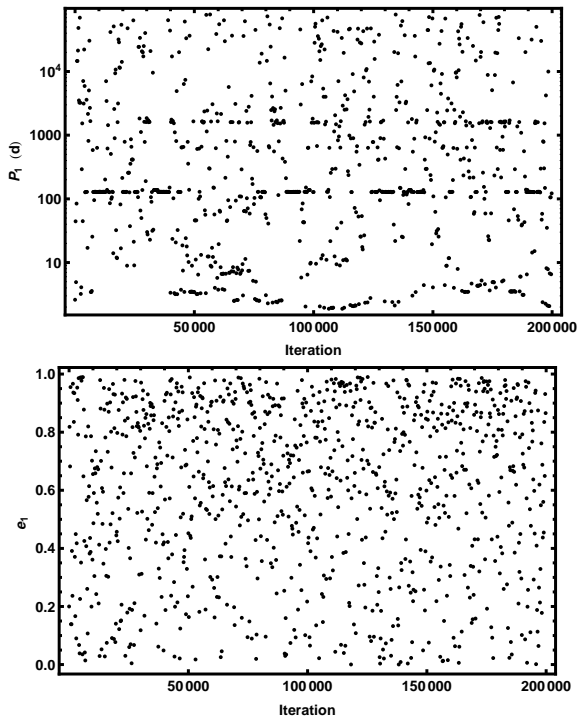}
\caption{The upper panel shows MCMC period parameter versus iteration for a 1 planet model fit to residuals (with randomized velocity values) from a 3 planet model fit. The lower panel is the same for the eccentricity parameter.}
\label{fig:randomV}
\end{figure}

We used two different choices of priors for eccentricity, a uniform prior and eccentricity noise bias correction filter that is described in the next section.

\subsection{Eccentricity Bias}
\label{sec:eccBias}

\begin{figure}
\includegraphics[width=85mm]{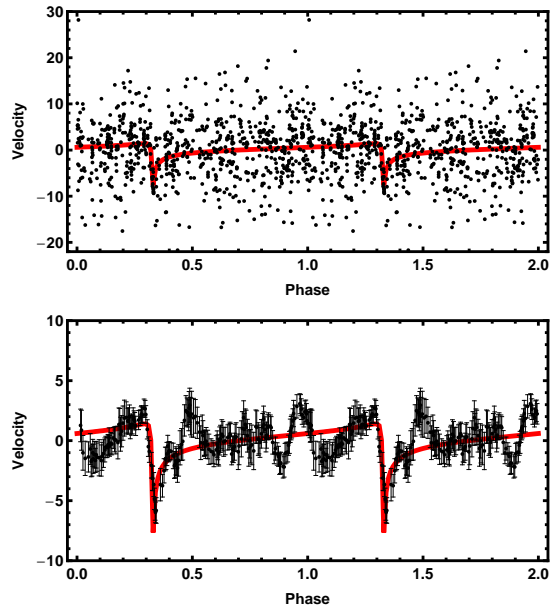}
\caption{A typical high eccentricity orbit (in this case $e = 0.93$) found from an MCMC fit of a 1 planet model to residuals with randomized velocities. The upper panel shows the raw data points plotted versus two cycles of period phase and the lower panel shows binned averages.}
\label{fig:typHigheccOrbit}
\end{figure}

When searching for low amplitude orbits any true signal has to compete against spurious orbital signals arising from noise. It was observed that the majority of the probability peaks detected in low signal-to-noise residuals exhibited high eccentricities. The upper panel in Figure~\ref{fig:randomV} shows MCMC period parameter versus iteration for a 1 planet model fit to residuals (with randomized velocity values) from a 3 planet model fit. The lower panel is the same for the eccentricity parameter. The HMCMC finds many probability peaks spread over the full period range. There is no significance to the concentration of periods around 100 and 1500 days as the location  of period concentrations changes markedly in other realizations of the velocity randomization. The concentration of eccentricity towards higher values is a regular feature. The corresponding plot of eccentricity shows a preponderance of high eccentricity values. Figure~\ref{fig:typHigheccOrbit} shows a phase plot for one of these high eccentricity orbits which provides further insight into why high eccentricities orbits are favored. It is clear that for most of the orbit ($e = 0.93$) the predicted shape is relatively flat providing an agreeable fit to points that fluctuate in an uncorrelated noise like fashion about some mean. Only for a small portion of the orbit does the noise have to conspire to give rise to the rapidly changing orbital velocity peak. To mimic a circular velocity orbit the noise points would have to appear correlated over a larger fraction of the orbit. For this reason it is more likely that noise will give rise to spurious highly eccentric orbits than low eccentricity orbits.  

\begin{figure}
\includegraphics[width=80mm]{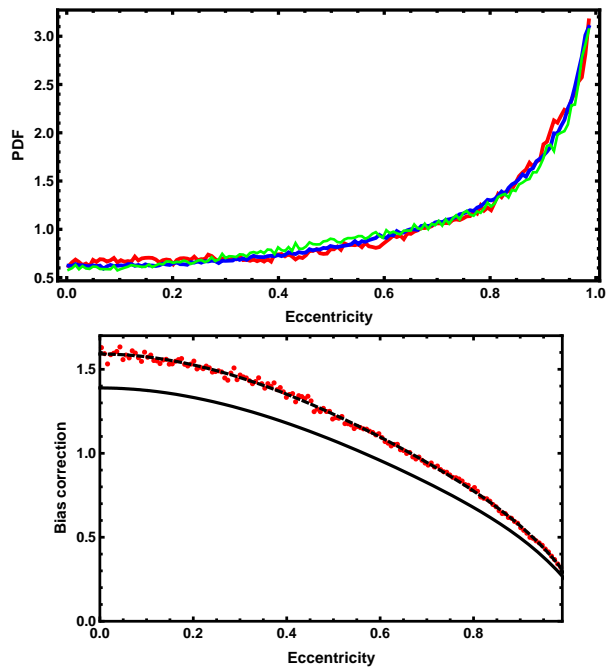}
\caption{The upper panel shows the marginal probability densities for the eccentricity parameter obtained from MCMC 1 planet fits to randomized residuals from 47 UMa 2 planet model fits (green), 3 planet (blue), and 55 Cancri 5 planet (red) fits. The green curve is the average of 4 trials, the blue curve is the average of 8 trials and the red curve is the average of 5 trials. The lower panel shows the best fit polynomial (dashed curve) to the reciprocal of the mean of the three eccentricity bias curves (red points). After normalization this yields the eccentricity noise bias correction filter (solid black curve).}
\label{fig:EccBias}
\end{figure}

To explore this effect more quantitatively we analyzed a large number of real data sets where the observing times were kept fixed but the velocity residual data was randomly reorganized. In each trial we fit a one planet orbit model which explored eccentricities in the range 0 to 0.99 using the one planet Bayesian Kepler periodogram. In the first instance the data used was the 5 planet fit residuals for 55 Cancri. The data for 55 Cancri was a mixture of Lick and Keck observatory data. When the residual velocities were randomized the error associated with a particular velocity was shifted with its velocity because the quoted errors were very different for the two observatories. The red curve in the left panel of Figure~\ref{fig:EccBias} is the average of 5 different 55 Cancri randomized residuals trials. The green curve is the average of 4 trials of randomized residuals from a 2 planet 47 UMa model fit, and the blue curve the average of 8 trials of randomized residuals from a 3 planet 47 UMa model fit. All three curves are very similar and indicate a strong noise induced eccentricity bias towards high eccentricities.  

To increase the chance of detecting and defining the parameters of low and moderate eccentricity orbits we have constructed an eccentricity noise bias correction filter from the reciprocal of the average of the three eccentricity bias curves just mentioned. The lower panel of Figure~\ref{fig:EccBias} shows the best fit polynomial (dashed curve) to the reciprocal of the mean of the three eccentricity bias curves (red points). After normalizing the best fit polynomial so the integral is equal to unity over the search range ($e =$ 0 to 0.99), we obtain the eccentricity noise bias correction filter (solid black curve). This becomes our second option for a choice of prior for eccentricity.
The probability density function for this filter (solid black curve) is given by
\begin{equation}
pdf(e) = 1.3889-1.5212 e^2 +0.53944 e^3 -1.6605(e-0.24821)^8.
\label{eq:EccDeBias}
\end{equation}
On the basis of our understanding of the mechanism underlying the noise induced eccentricity bias, we expect the eccentricity prior filter to be generally applicable to searches for low amplitude orbital signals in other precision radial velocity data sets.

An obvious further question that remains to be explored is to what extent the observed distribution of published orbital eccentricities is influenced by such a bias.

\section{Results (Case A)}

For Case A, the dewar velocity offsets with respect to our reference dewar \# 24 are assumed to be zero.

\subsection{Parameter estimation}
\label{sec:ParEstimate}

\begin{figure}
\includegraphics[width=80mm]{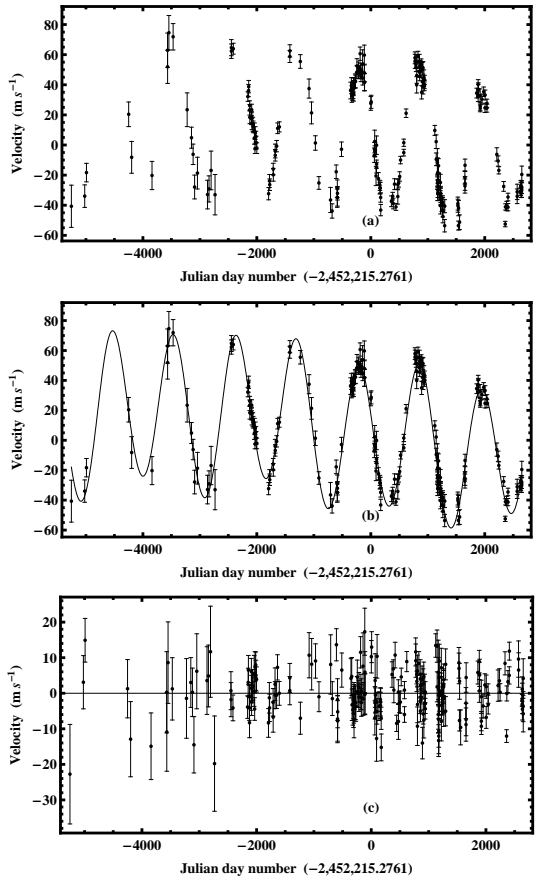}
 \caption{Panel (a) shows the Lick Observatory observations of 47 UMa. Panel (b) shows the final Case A best 3 planet model fit compared to the data and panel (c) shows the residuals.}
\label{fig:47UMa_data}
\end{figure}

In this section we present the results of an exploration of the 47 UMa data with the multi-planet HMCMC Kepler periodogram starting with a one planet model and extending to a five planet model. The data for 47UMa is shown in Figure~\ref{fig:47UMa_data} panel (a). Panel (b) shows our final best 3 planet model fit compared to the data and panel (c) shows the residuals. 

The one planet model turned up the 1080 day period which is clearly visible by eye in the raw data. We do not show any results for that model except to compute the marginal likelihood for model selection purposes which is presented in Section~\ref{sec:modsel}.

\begin{figure}
\includegraphics[width=80mm]{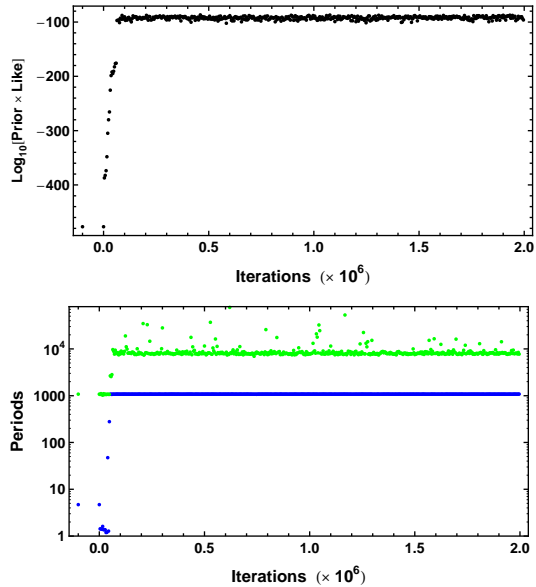}
\caption{Plot of Log$_{10}$[Prior $\times$ Likelihood] (upper) and period (lower) versus MCMC iteration for a 2 planet model.}
\label{fig:2planProbPvsIter}
\end{figure}

Figure~\ref{fig:2planProbPvsIter} shows a plot of Log$_{10}$[Prior $\times$ Likelihood] (upper) and period (lower) versus HMCMC iteration (every 200$^{\rm th}$ point) for a 2 planet model. The starting periods of 4.7 and 1080 days are shown on the left hand side of the lower plot at a negative iteration number. The burn-in period of approximately 70,000 iterations is clearly discernable.

\begin{figure}
\begin{center}
\includegraphics[width=70mm]{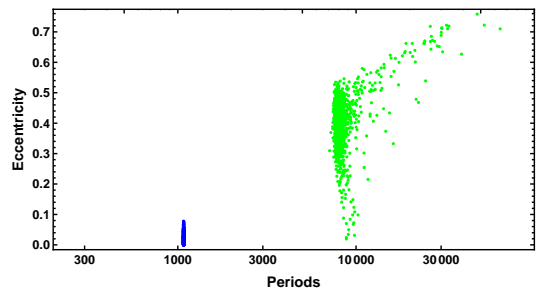}
\end{center}
 \caption{A plot of eccentricity versus period for the 2 planet fit (Case A).}
\label{fig:2planEccP}
\end{figure}

Figure~\ref{fig:2planEccP} shows a plot of eccentricity versus period for a sample of the HMCMC parameter samples for the 2 planet model. Since the duration of the data set is only 7906 days, it is not surprising that uncertainties on the parameters of the second orbit are very large.
On the basis of a 2 planet model, the parameters of the second planet are $P_2 = 7952_{-348}^{+388}$d and $e_2 = 0.43_{-0.08}^{+0.05}$. It is clear that $e_2$ has a low eccentricty tail which reaches zero for a value of $P_2 \approx 9500$d. This agrees with the value of $P_2 = 9660$d found by \citet{Wittenmyer2009} in their best-fit 2-planet model where they fixed $e_2= 0.005$, the values proposed by \citet{Fischer2002}. 

\begin{figure}
\includegraphics[width=80mm]{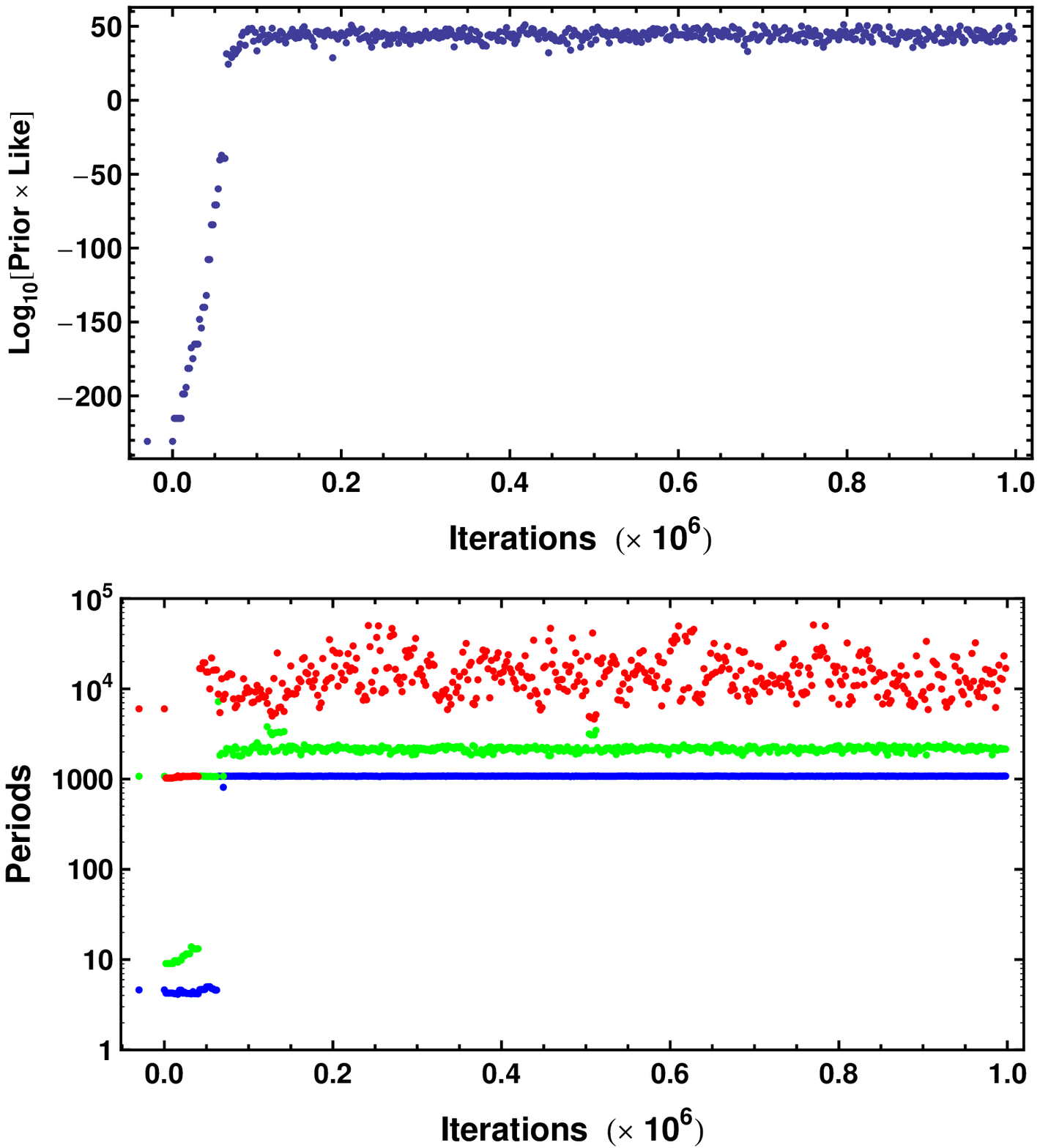}
\caption{Plot of Log$_{10}$[Prior $\times$ Likelihood] (upper) and period (lower) versus HMCMC iteration for a 3 planet fit.}
\label{fig:3planProbPvsIter}
\end{figure}

\begin{figure}
\includegraphics[width=85mm]{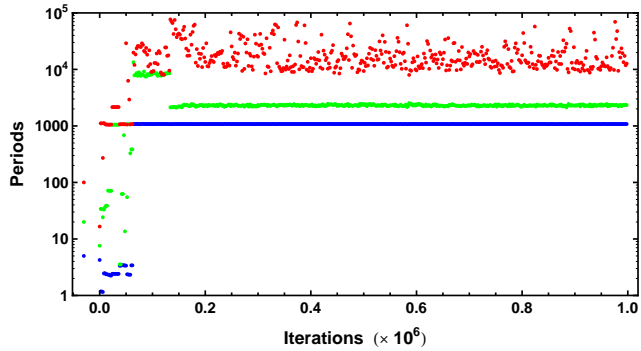}
\caption{Plot of period versus HMCMC iteration for a 3 planet fit. In this case the start periods were 5, 20, 100 days.}
\label{fig:3planPvsIterC}
\end{figure}

\begin{figure}
\includegraphics[width=80mm]{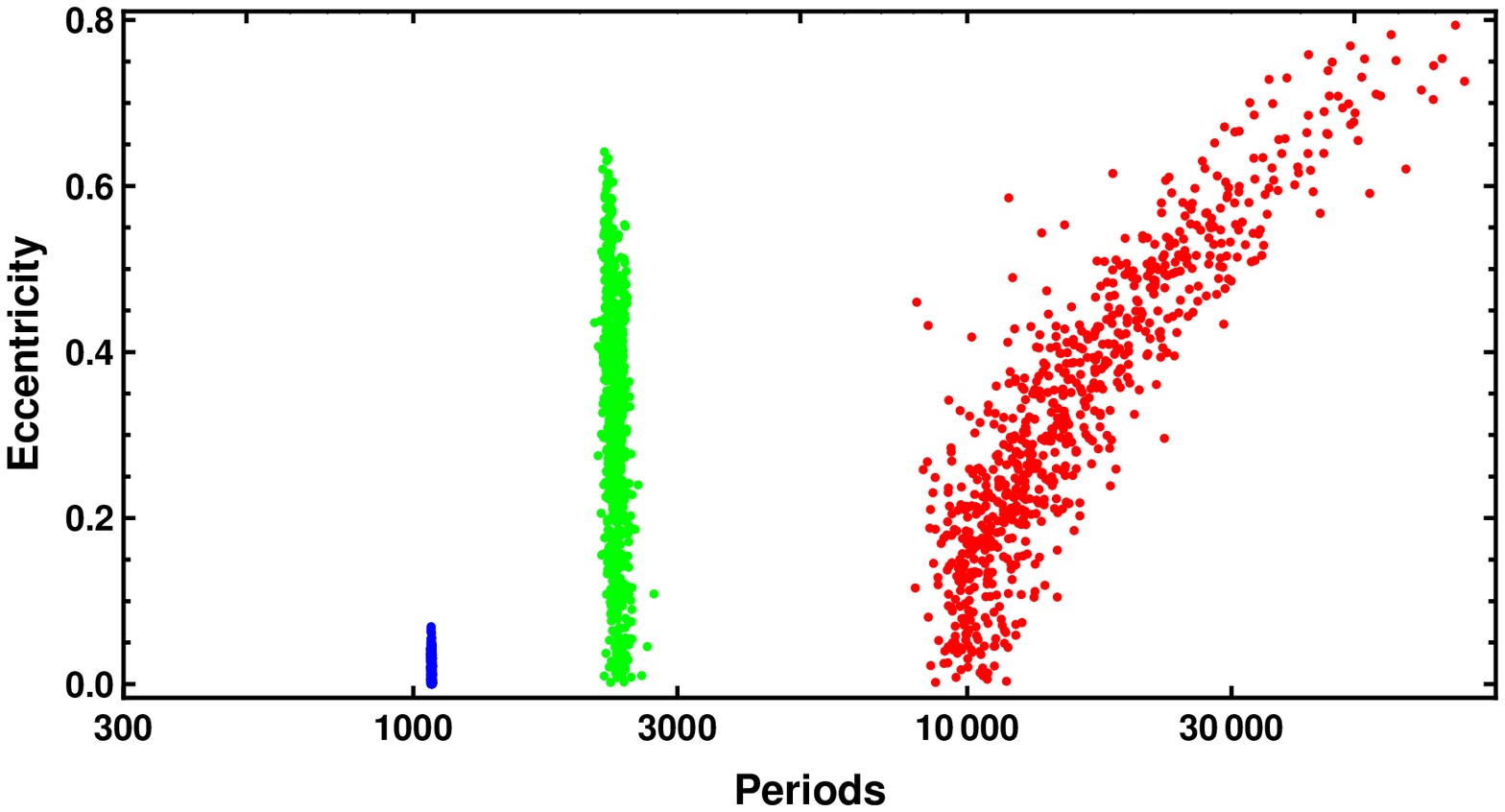}
\caption{A plot of eccentricity versus period for the 3 planet HMCMC (Case A).}
\label{fig:3planEccP}
\end{figure}

Figure~\ref{fig:3planProbPvsIter} shows plots of the 3 period parameters versus HMCMC iteration for a 3 planet model~\footnote{Note: the HMCMC runs shown here used the eccentricity prior based on the eccentricity noise bias correction filter discussed in Section~\ref{sec:eccBias}. The results obtained using a uniform eccentricity prior are qualitatively the same.} with Log$_{10}$[Prior $\times$ Likelihood] plotted above. A new period of 2300d has emerged and the longest period has shifted from 7952d to $\sim 10000$d and this feature is considerably broader. The starting periods of 89, 1080, 7200d are shown on the left at a negative iteration number. Previous experience with the HMCMC periodogram (\citealt{Gregory2009}) indicate that it is capable of finding a global peak in a blind search of parameter space for a three planet model. Figure~\ref{fig:3planPvsIterC} shows the results of a blind search starting from 3 very different periods of 5, 20, 100d. The algorithm readily finds the same set of final periods in both cases.  
Figure~\ref{fig:3planEccP} shows a plot of eccentricity versus period for a sample of the HMCMC parameter samples for the 3 planet model. There is a large uncertainty in the eccentricity of the two largest periods which extends down to very low eccentricities.

\begin{figure*}
\includegraphics[width=160mm]{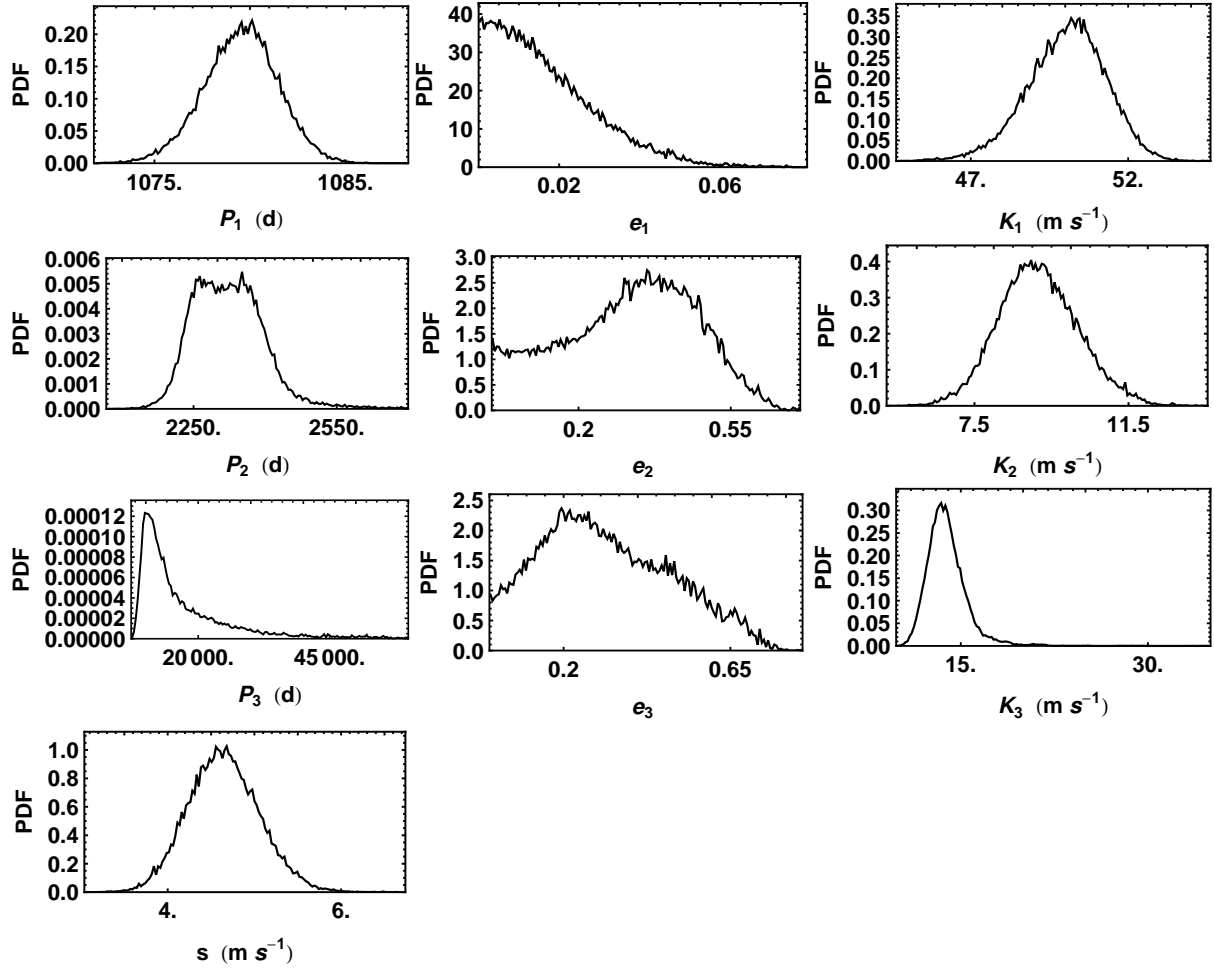}
\caption{A plot of parameter marginal distributions for a 3 planet HMCMC (Case A).}
\label{fig:3planMarg}
\end{figure*}

Figure~\ref{fig:3planMarg} shows the marginal probability distributions for the periods, eccentricities and K values for the three orbits found. The tenth plot is $s$, the $\sigma$ of the added white noise term. A summary of the 3 planet model parameters and their uncertainties are given in Table~\ref{tab:parerrorsM3}. The parameter value listed is the median of the marginal probability distribution for the parameter in question and the error bars identify the boundaries of the 68.3\% credible region~\footnote{In practice, the probability density for any parameter is represented by a finite list of values $p_i$ representing the probability in discrete intervals $X$. A simple way to compute the 68.3\% credible region, in the case of a marginal with a single peak, is to sort the $p_i$ values in descending order and then sum the values until they approximate 68\%, keeping track of the upper and lower boundaries of this region as the summation proceeds.}. The value immediately below in parenthesis is the maximum {\it a posteriori} (MAP) value, the value at the maximum of the joint posterior probability distribution. It is not uncommon for the MAP value to fall close to the borders of the credible region. In one case, the period of the third planet, the MAP value falls outside the 68.3\% credible region which is one reason why we prefer to quote median values as well. The marginal for $P_3$ is so asymmetric we also give the mode which is 9991 days. The semi-major axis and $M \sin i$ values are derived from the model parameters assuming a stellar mass of $1.063_{+0.029}^{-0.022}$ M$_{\sun}$ \citep{Takeda2007}. The quoted errors on the semi-major axis and $M \sin i$ include the uncertainty in the stellar mass. 

\begin{table}
 \centering
 \begin{minipage}{140mm}
  \caption{Three planet model parameter estimates (Case A).}
  \label{tab:parerrorsM3}
  \begin{tabular}{@{}lllll@{}}
  \hline
   Parameter  & planet 1 & planet 2 & planet 3 \\
\hline
$P$  (d) & $1079.6_{-1.8}^{+2.0}$ & $2319_{-76}^{+63}$& $13346_{-4940}^{+4030}$  \\
& (1079.2)& (2278)& (21342) \\
& & &mode$=9991$  \\
& & &  \\
$K$ (m s$^{-1}$) & $50.1_{-1.2}^{+1.3}$ & $9.1_{-1.0}^{+1.0}$ & $13.7_{-1.4}^{+1.3}$  \\
& (50.3) & (9.6) & (13.2) \\
& & &  \\
$e$ & $0.014_{-.014}^{+.008}$ & $0.33_{-.17}^{+.2}$  & $0.29_{-.21}^{+.21}$   \\
& (0.012) & (0.48) & (0.44) \\
& & &  \\
$\omega$  (deg) & $350_{-69}^{+84}$ & $222_{-21}^{+21}$ &  $162_{-50}^{+40}$  \\
& (345) & (222) & (111) \\
& & &  \\
$a$  (au) & $2.10_{-.02}^{+.02}$ & $3.50_{-.08}^{+.07}$ &  $11.3_{-2.8}^{+2.2}$  \\
& (2.10) & (3.46) &  (15.4)  \\

& & & & \\
$M \sin i$  ($M_J$) & $2.63_{-.07}^{+.09}$ & $0.575_{-.056}^{+.052}$ &  $1.58_{-.18}^{+.17}$ \\
& (2.64) & (0.566) &  (1.69) \\
& & &  \\
Periastron & $11967_{-202}^{+252}$ & $11914.6_{-131}^{+166}$ &  $12655_{-4543}^{+5144}$  \\
\ passage &  (11943) & (11930)&  (12047) \\
\ (JD - 2,440,000) & & &  \\
\hline
\end{tabular}
\end{minipage}
\end{table}

The \citet{Gel} statistic is typically used to test for convergence of the parameter distributions. In parallel tempering MCMC, new widely separated parameter values are passed up the line to the $\beta = 1$ simulation and are occasionally accepted. Roughly every 100 iterations the $\beta = 1$ simulation accepts a swap proposal from its neighboring simulation. 
The final $\beta = 1$ simulation is thus an average of a very large number of independent $\beta = 1$ simulations. What we have done is divide the $\beta = 1$ iterations into 12 equal time intervals and inter-compared the 12 different essentially independent average distributions for each parameter using a Gelman-Rubin test. For all of the three planet model parameters the Gelman-Rubin statistic was $\le 1.07$.

\begin{figure}
\includegraphics[width=85mm]{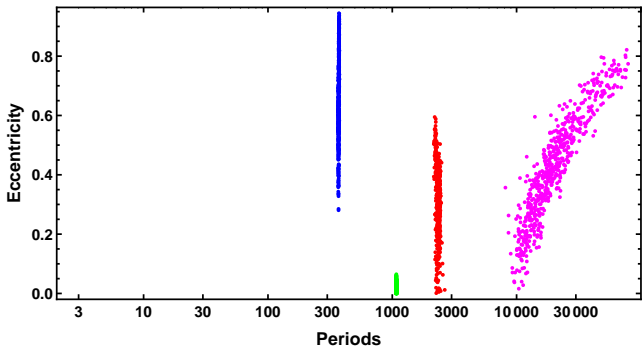}
 \caption{A plot of eccentricity versus period for the 4 planet HMCMC (Case A).}
\label{fig:4planEccP}
\end{figure}

Figure~\ref{fig:4planEccP} shows a plot of eccentricity versus period for a 4 planet model. A well defined fourth period of $370.8_{-2.0}^{+2.4}$ days and eccentricity of $0.57_{-0.15}^{+0.22}$ was detected in repeated HMCMC trials.   The amplitude was $K = 5.0_{-1.1}^{+1.0}$m s$^{-1}$. The significance of this period is discussed further in Sections~\ref{sec:modsel} and \ref{sec:discussion}.

Finally, a 5 planet model was also attempted. In addition to the 4 periods found by the 4 planet model, a variety of probability peaks at other periods were observed but none were deemed significant.

\subsubsection{Simulation test}
\label{sec:SimTest}

\begin{figure}
\includegraphics[width=80mm]{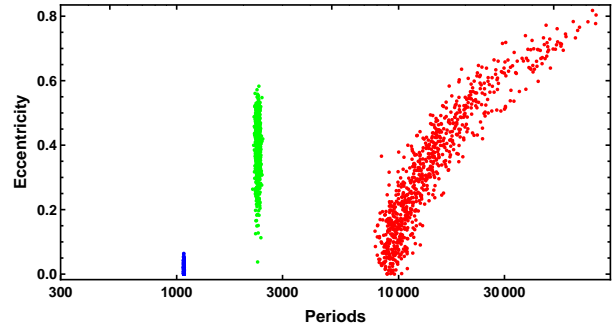}
\caption{A plot of eccentricity versus period for a 3 planet HMCMC fit of the 3 planet simulation.}
\label{fig:3planEccPSim}
\end{figure}
As a test of our overall methodology, we simulated data for a 3 planet model based on the MAP values from the fit to the real data for the Case A analysis. The data was sampled at the real observation times and had added independent Gaussian noise with a $\sigma = \sqrt{(e_i)^2 + s^2}$, where $e_i$ is the quoted measurement error for the i$^{th}$ point and $s$, the extra noise parameter, was 4.4m s$^{-1}$. 
Figure~\ref{fig:3planEccPSim} shows a plot of eccentricity versus period for a sample of the HMCMC parameter samples for the 3 planet model fit to the simulated data set. Again, the starting period values for the HMCMC were 5, 20, and 100 days, a long way from the expected values. Comparison with Figure~\ref{fig:3planEccP} indicates that the results for the actual data and 3 planet simulation are qualitatively very similar.

To test whether the fourth period in the Lick data (period = $370.8_{-2.0}^{2.4}$ days) is a window function artifact of the sampling times, we analyzed two 3 planet simulations with a 4 planet model. In both cases the HMCMC found the 3 periods expected from the simulation. No well defined fourth period was found and the peak amplitude was $K = 3$m s$^{-1}$ compared with a $K = 5$m s$^{-1}$ for the real data set. This suggests that the fourth period is not simply a window function artifact. However, later HMCMC fits of a combination of Lick and Mcdonald Observatory data did not confirm this period.

\subsection{Model selection}
\label{sec:modsel}

On of the great strengths of Bayesian analysis is the built-in Occam's razor. More complicated models contain larger numbers of parameters and thus incur a larger Occam penalty, which is automatically incorporated in a Bayesian model selection analysis in a quantitative fashion (see for example, \citealt{Gregorybook}, p. 45). The analysis yields the relative probability of each of the models explored.

To compare the posterior probabilities of the i$^{\rm th}$ planet model to the one planet models we need to evaluate the odds ratio,
$O_{i1} =p(M_{i} | D,I)/p(M_{1} | D,I)$, the ratio of the posterior probability
of model $M_{i}$ to model $M_{1}$.  Application of Bayes's 
theorem leads to,
\begin{equation}
O_{i2} = {p(M_{i} | I) \over p(M_{1} | I)}\;
      {p(D | M_{i},I) \over p(D | M_{1},I)}
       \equiv {p(M_{i} | I) \over p(M_{1} | I)}\; B_{i2}
\label{eq:orbit22}
\end{equation}
where the first factor is the prior odds ratio, and the second factor
is called the {\it Bayes factor}, $B_{i2}$. The Bayes factor is the ratio of
the marginal (global) likelihoods of the models. The marginal likelihood for model $M_i$ is given by
\begin{equation}
p(D|M_i,I)= \int d\vec{X} p(\vec{X}|M_i,I)\times p(D|\vec{X},M_i,I).
\label{eq:marglike}
\end{equation}
Thus Bayesian model selection relies on the ratio of marginal likelihoods, not maximum likelihoods. The marginal likelihood is the weighted average of the conditional likelihood, weighted by the prior probability distribution of the model parameters and $s$. This procedure is referred to as marginalization.  

The marginal likelihood can be expressed as the product of the maximum likelihood and the Occam penalty (see \citealt{GregoryLoredo1992} and \citealt{Gregorybook}, page 48). The Bayes factor will favor the more complicated model only if the maximum likelihood ratio is large enough to overcome this penalty. In the simple case of a single parameter with a uniform prior of width $\Delta X$, and a centrally peaked likelihood function with characteristic width $\delta X$, the Occam factor is $\approx \delta X/\Delta X$. If the data is useful then generally $\delta X \ll \Delta X$. For a model with $m$ parameters, each parameter will contribute a term to the overall Occam penalty. The Occam penalty depends not only on the number of parameters but also on the prior range of each parameter (prior to the current data set, $D$), as symbolized in this simplified discussion by $\Delta X$. If two models have some parameters in common then the prior ranges for these parameters will cancel in the calculation of the Bayes factor. To make good use of Bayesian model selection, we need to fully specify priors that are independent of the current data $D$. The sensitivity of the marginal likelihood to the prior range depends on the shape of the prior and is much greater for a uniform prior than a Jeffreys prior (e.g., see \citealt{Gregorybook}, page 61). In most instances we are not particularly interested in the Occam factor itself, but only in the relative probabilities of the competing models as expressed by the Bayes factors. Because the Occam factor arises automatically in the marginalization procedure, its effect will be present in any model selection calculation. Note: no Occam factors arise in parameter estimation problems. Parameter estimation can be viewed as model selection where the competing models have the same complexity so the Occam penalties are identical and cancel out. 
   
The MCMC algorithm produces samples which are in proportion to the posterior probability distribution which is fine for parameter
estimation but one needs the proportionality constant for estimating the model marginal likelihood. 
\citet{Clyde2006} recently reviewed the state of techniques for model selection from a statistics perspective and \citet{FordGregory2006} have evaluated the performance of a variety of marginal likelihood estimators in the extrasolar planet context. 

\citet{Gregory2007c}, in the analysis of velocity data for HD 11964, compared the results from three marginal likelihood estimators: (a) parallel tempering, (b) ratio estimator, and (c) restricted Monte Carlo. Monte Carlo (MC) integration can be very inefficient in exploring the whole prior parameter range because it randomly samples the whole volume. The fraction of the prior volume of parameter space containing significant probability rapidly declines as the number of dimensions increases. For example, if the fractional volume with significant probability is 0.1 in one dimension then in 17 dimensions the fraction might be of order $10^{-17}$. In restricted MC integration (RMC) this should be much less of a problem because the volume of parameter space sampled is restricted to a region delineated by the outer borders of the marginal distributions of the parameters. 
For HD 11964, the three methods were compared for 1, 2 and 3 planet models. For the one planet model all three methods agreed within 15\%. For the two planet model the three methods agreed within 28\% with the RMC giving the lowest estimate. For the three planet model the estimates were very different. The RMC estimate was 16 time smaller than the PT estimate and the ratio estimator was 18 times larger than the PT estimate. The PT method is very compute intensive. For a three planet model 40 tempering levels and $10^7$ iterations were required. The problem becomes more difficult for larger numbers of planets. Thus for 3 or more planet models accurately computing the marginal likelihood is a very big challenge. 

\begin{table*}
  \caption{Marginal likelihood estimates, Bayes factors and false alarm probabilities for (Case A) 0, 1, 2, 3, and 4 planet models which are designated $M_0, \cdots, M_4$. The last two columns list the MAP value of extra noise parameter, $s$, and the RMS residual.}
  \label{tab:modelSel}
  \begin{tabular}{@{}llllllll@{}}
  \hline
   Model & Periods &  Marginal & Bayes factor & False Alarm & $s$ &RMS residual \\
         & (d) &  Likelihood & \ \ nominal & Probability &  (m s$^{-1})$ & (m s$^{-1}$)\\
\hline
$M_{0}$ & & $ 2.63\times 10^{-481}$ & $10^{-127}$ & & 34.8 & 35.3\\
& & & & & &\\
$M_{1}$ & $(1080)$& $(7.51\pm 0.07)\times 10^{-394}$ & $10^{-39}$ & $10^{-88}$ & 11.2 & 12.5\\
& & & & & &\\
$M_{2}$  & $(1080,8000)$& $(4.1\pm 0.5)\times 10^{-355}$ & 1.0 & $10^{-39}$ & 6.1 & 8.1\\
& & & & & &\\
$M_{3}$ & $(1080,2300,\sim10000)$ & $(4^{\times 2}_{\times 1/5}) \times 10^{-338}$ & $10^{17}$ & $10^{-17}$ & 4.4 & 6.5\\
& & & & & &\\
$M_{4}$  &$(371,1080,2300,\sim10000)$& $(4^{\times 7}_{\times 1/2}) \times 10^{-338}$ & $10^{17}$ & $0.5$ & 3.7 & 6.1\\
\hline
\end{tabular}
\end{table*}
 
In this work we consider only RMC marginal likelihood estimates. This method is expected to underestimate the marginal likelihood in higher dimensions and this underestimate is expected to become worse the larger the number of model parameters, i.e. increasing number of planets. When we conclude, as we do, that the RMC computed odds in favor of the three planet model compared to the two planet model is $\sim 10^{17}$ we mean that the true odds is $\ge 10^{17}$. 

In earlier work, we defined the outer boundary of parameter space for RMC integration based on the 99\% credible region. One problem is that if there is a significant contribution to the integral within say the 30\% credible region, the volume in this region can be such a small fraction of the total that no random sample lands in that region. In this work we use a nested version of RMC integration. Multiple boundaries were constructed based on credible regions ranging from 30\% to $\ge 99\%$, as needed. We are then able to compute the contribution to the total RMC integral from each nested interval and sum these contributions. For example, for the interval between the 30\% and 60\% credible regions, we generate random parameter samples within the 60\% region and reject any sample that falls within the 30\% region. Using the remaining samples we can compute the contribution to the RMC integral from that interval. 

\begin{figure*}
\includegraphics[width=160mm]{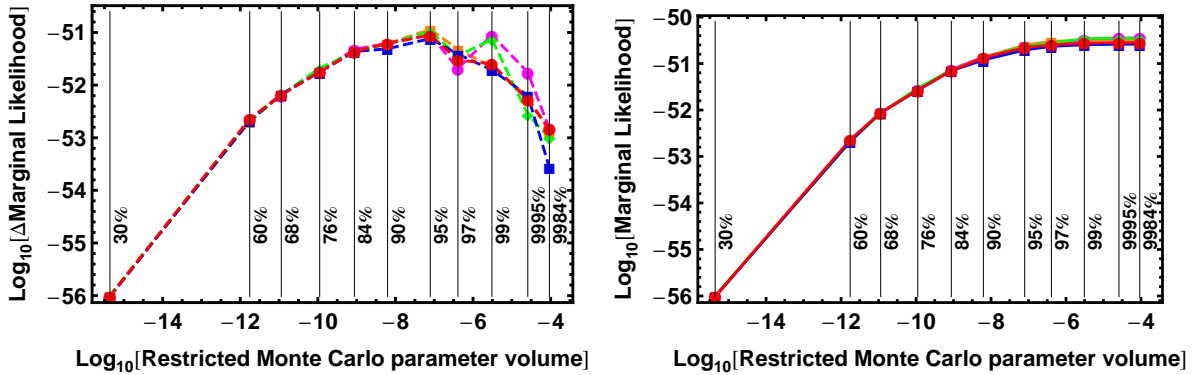}
\caption{Left panel shows the contribution of the individual nested intervals to the RMC marginal likelihood for the 2 planet model. The right panel shows the integral of these contributions versus the parameter volume of the credible region. Note: only the relative values of the units on the vertical axes of these two plots are meaningful.}
\label{fig:2planNestedRMC}
\end{figure*}

The left panel of Figure~\ref{fig:2planNestedRMC} shows the contributions from the individual intervals for 5 repeats of the nested RMC evaluation for the 2 planet model. The right panel shows the summation of the individual contributions versus the volume of the credible region. The credible region listed as 9995\% is defined as follows. Let $X_{U99}$ and $X_{L99}$ correspond to the upper and lower  boundaries of the 99\% credible region, respectively, for any of the parameters. Similarly, $X_{U95}$ and $X_{L95}$ are the upper and lower boundaries of the 95\% credible region for the parameter. Then $X_{U9995} = X_{U99}+(X_{U99}-X_{U95})$ and $X_{L9995} = X_{L99}+(X_{L99}-X_{L95})$. Similarly, $X_{U9984} = X_{U99}+(X_{U99}-X_{U84})$. 

Table~\ref{tab:modelSel} gives the Marginal likelihood estimates, Bayes factors and false alarm probabilities for 0, 1, 2, 3, and 4 planet models which are designated $M_0, \cdots, M_4$. The last two columns list the MAP value of extra noise parameter, $s$, and the RMS residual. For each model the RMC calculation was repeated 5 times and the quoted errors give the spread in the results, not the standard deviation. The Bayes factors that appear in the third column are all calculated relative to model 2. Examination of a plot like the one shown in Figure~\ref{fig:2planNestedRMC}, but for the 4 planet model, indicates that RMC is probably seriously underestimating the marginal likelihood. A better method of computing this quantity is sorely needed. 

We can readily convert the Bayes factors to a Bayesian False Alarm Probability (FAP). For example, in the context of claiming the detection of a 3 planet model the FAP is the probability that there are actually 2 or less planets.
\begin{equation}
{\rm FAP} = \sum_{i=0}^{2}({\rm prob. of \ i \ planets}) 
\label{eq:FAP1}
\end{equation}

If we assume {\it a priori} (absence of the data) that the prob of 1 planet model = prob. of 2 planet model --- etc., then probability of each model is related to the Bayes factors by
\begin{equation}
p(M_i\mid D,I) = {{B_{i2}} \over {\sum_{j=0}^{N_{\rm mod}} B_{j2}}}
\label{eq:FAP2}
\end{equation}
where $N_{\rm mod}$ is the total number of models considered, and of course
$B_{22} = 1$. Given the Bayes factors in Table~\ref{tab:modelSel} and substituting into equation~\ref{eq:FAP1} gives
\begin{equation}
{\rm FAP} = {{(B_{02} + B_{12} + B_{22})}\over {\sum_{j=0}^{3} B_{j2}}}\approx 10^{-17}
\label{eq:FAP3}
\end{equation}
For the 3 planet model we obtain a very low FAP $\approx 10^{-17}$. The Bayesian false alarm probabilities for 1, 2, 3, and 4 planet models are given in the fourth column of Table~\ref{tab:modelSel}.

In the context of claiming the detection of a 4 planet model the Bayesian false alarm probability is $\approx 0.5$. This is very high and does not justify a claim for the detection of a fourth planet. The fourth period is also suspiciously close to one year to be of concern.

\section{Results (Case B)}

For Case B, we incorporated 4 additional parameters to allow for the unknown residual velocity offsets of dewars 6, 8, 39, and 18 relative to dewar 24. These are labeled $V_{6}, V_{8}, V_{39}, V_{18}$, where the subscript denotes the detector dewar. In a Bayesian analysis these are treated as additional nuisance parameters which we can marginalize. Additionally, since they are of interest to the observers we also provide a summary of each residual offset parameter. In the radial velocity data processing pipeline every effort was made to insure the dewar velocity offsets were allowed for so the residuals are expected to be small. For the Case B analysis we have assumed a Gaussian prior for each $V_{i}$ centered on zero with a $\sigma = 3$ km s$^{-1}$.

\subsection{Parameter estimation}
\label{sec:ParEstimate}

In this section we redo the analysis of the 47 UMa data with the multi-planet HMCMC Kepler periodogram starting with a one planet model and extending to a four planet model. The data is same as shown in Figure~\ref{fig:47UMa_data} panel (a) with the exception of the first point corresponding to dewar 1.  

\begin{figure}
\begin{center}
\includegraphics[width=70mm]{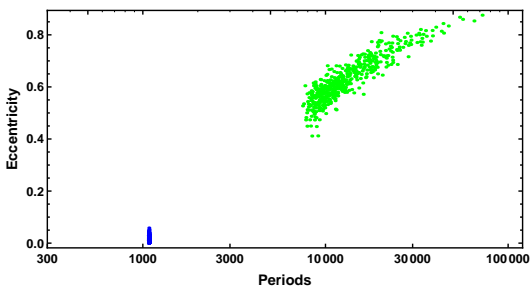}
\end{center}
 \caption{A plot of eccentricity versus period for the 2 planet fit (Case B).}
\label{figB:2planEccP}
\end{figure}

Figure~\ref{figB:2planEccP} shows a plot of eccentricity versus period for a sample of the HMCMC parameter samples for the 2 planet model for Case B. The two planet model again favors a second period in the range 8100-15000d (68\% credible region) with a long higher eccentricity tail extending to much longer periods. In Case B the time base is 235 days shorter than Case A so the lower eccentricity/lower period end is less well defined but otherwise there is general agreement. This model was run twice using different starting periods but the two planet HMCMC run did not favor a period around 2240 days even when the two starting periods used were 1078 and 2240 days, respectively. This is not that surprising given the relative sizes of the $K$ values for planets 2 and 3 in Table~\ref{tab:parerrorsM3}.

\begin{figure}
\includegraphics[width=80mm]{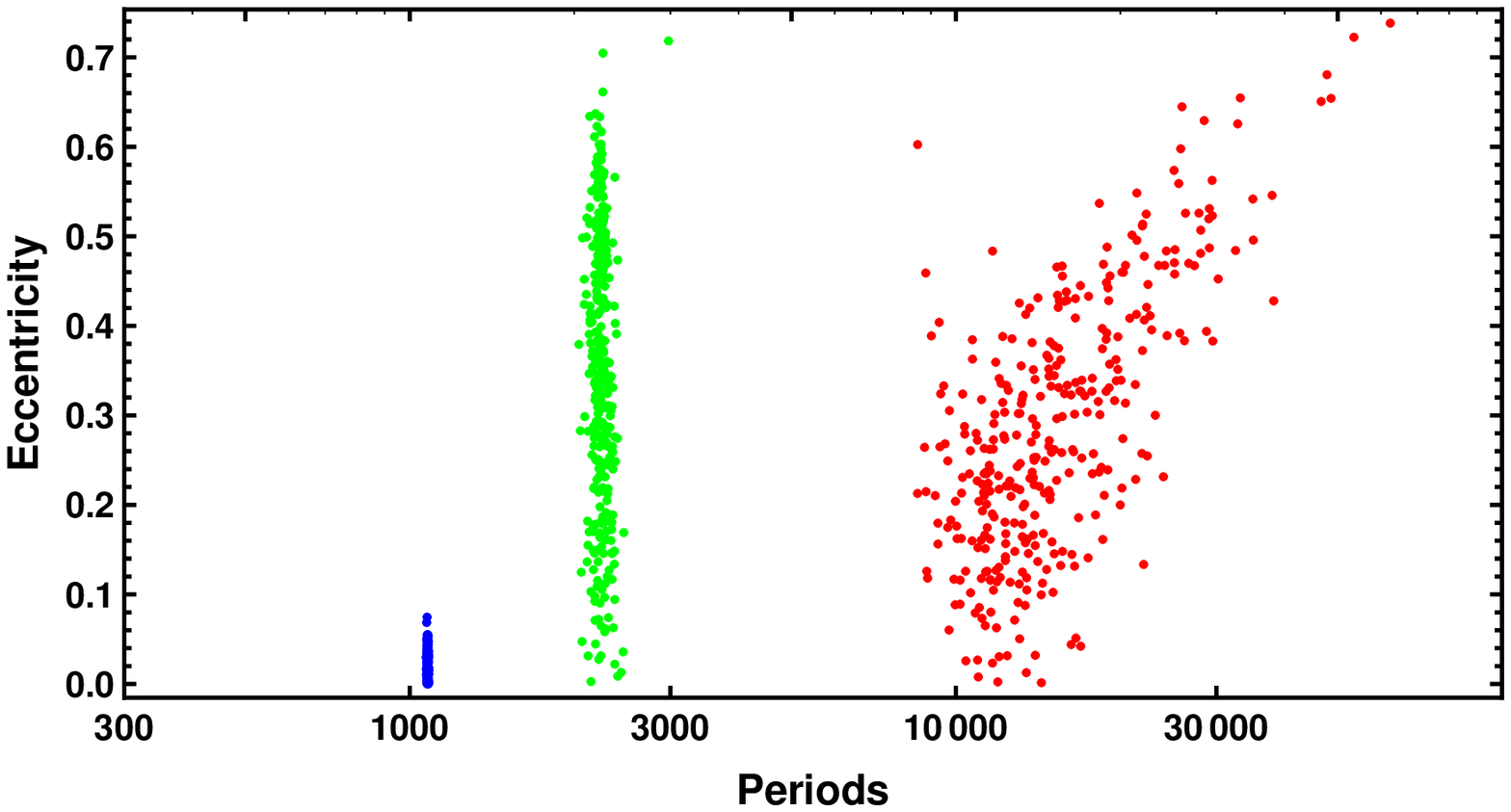}
\caption{A plot of eccentricity versus period for the 3 planet HMCMC (Case B).}
\label{figB:3planEccP}
\end{figure}

Figure~\ref{figB:3planEccP} shows a plot of eccentricity versus period for a sample of the HMCMC parameter samples for the 3 planet model for Case B. Again, we see the emergence of a period of $\sim 2250$ days and the third longer period appears better defined (compared to the 2 planet model) and extends to much lower eccentricities. Qualitatively, there is general agreement with the Case A results shown in Figure~\ref{fig:3planEccP}.
The dewar residual offset velocities were $V_{6}$ = $0.07_{-2.6}^{+2.7}$,  $V_{8}$ = $1.7_{-2.3}^{+3.0}$, $V_{39}$ = $-3.2_{-2.4}^{+2.5}$, and $V_{18}$ =  $-1.1_{-2.0}^{+2.0}$ m s$^{-1}$. 

The 4 planet HMCMC analysis again showed a clear fourth period of $372_{-1.3}^{1.9}$ with an eccentricity of $0.73\pm 0.14$. We did not compute the marginal likelihood for the 4 planet model but based on the Case A results the Bayesian false alarm probability for a 4 planet model is expected to be very high. 

\subsection{Model selection (Case B)}
\label{sec:modselB}

We repeated the Bayesian false alarm probability for the 3 planet model as described in Section~\ref{sec:modsel} for the Case B analysis which incorporates the dewar residual offset parameters.
\begin{equation}
{\rm FAP} = {{(B_{02} + B_{12} + B_{22})}\over {\sum_{j=0}^{3} B_{j2}}}
\label{eq:FAP3}
\end{equation}
The computed Bayes factors are $B_{02} = 1.6 \times 10^{-141} , B_{12} = 2.0 \times 10^{-28} , B_{22} = 1.0, B_{32} = 2.0 \times 10^5$. This gives a false alarm probability of $5.0 \times 10^{-6}$. Even though this is much greater than the value found in Case A it still argues strongly for favoring a 3 planet model. 

\section{Discussion}
\label{sec:discussion}

On the basis of the model selection results we can conclude there is strong evidence for three planets although the longest period orbital parameters are still not well defined. The results for the Lick only analysis do not rule out low eccentricity orbits for all three planets. The major difference produced by including the dewar residual offset parameters was to reduce the Bayesian false alarm probability for a 3 planet model from $\sim 10^{-17}$ to  to $\sim 10^{-5}$. A significant part of this reduction might be a consequence of the reduced span of the data set by 235 days for the Case B analysis.

Our results appear to be entirely consistent with the latest analysis of \citet{Wittenmyer2009}. Their best-fit 2-planet model now calls for $P_2 = 9660$ days. They note that to fit a second planet, they fixed the parameters $e_2$ and $\omega_2$ at the values proposed by \citet{Fischer2002}: e2 = 0.005 and $\omega_2 =127$. In our Case A two planet fit (Figure~\ref{fig:2planEccP}), in which all parameters were free, the eccentricity versus period plot exhibits a low eccentricity tail which occurs at a period between $9000$ and $10000$ days, directly comparable to their 9660 day period. The $\sim 2300$ day period in the Lick data only shows up in our 3 planet and higher models. This is probably because the longer period signal with a $K = 13.8$ m s$^{-1}$ dominates over the $2300$ day period signal with a $K = 8.0$ m s$^{-1}$ (see Table~\ref{tab:parerrorsM3Combined}). \citet{Wittenmyer2009} did not report any results on fitting a 3 planet model.
 
To test this further we combined the Lick data with the \citet{Wittenmyer2009} data from the 9.2 m Hobby-Eberly Telescope (HET) and 2.7 m Harlam J. Smith (HJS) telescopes of the McDonald Observatory. We subtracted initial offset velocities of 23.3 and 25.4 m s$^{-1}$ based on a comparison of plots of the HET and HJS data sets to the Lick data. We then included a free parameter for each telescope to allow for an unknown residual velocity offset compared with the Lick dewar 24 in the same way as we had done for the other Lick dewars in Case B. 

\begin{figure}
\includegraphics[width=80mm]{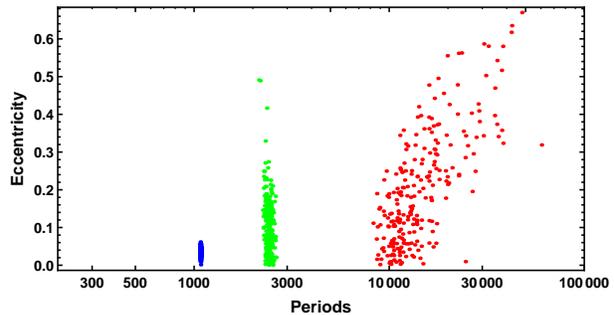}
\caption{A plot of eccentricity versus period for a 3 planet HMCMC fit of the combined Lick, HET, and HJS telescope data set.}
\label{fig:3planLHetEccP}
\end{figure}

\begin{figure*}
\includegraphics[width=140mm]{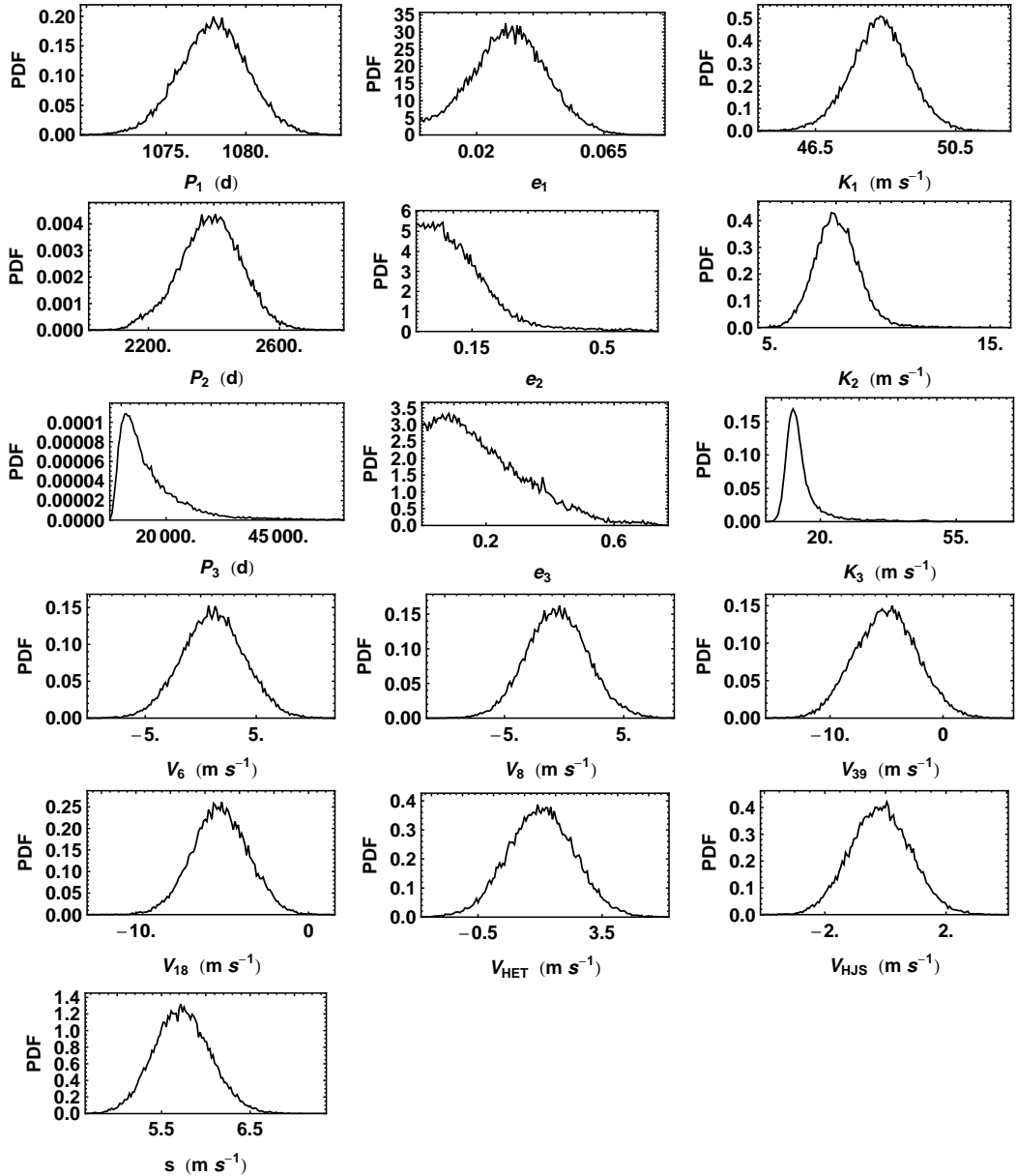}
\caption{A plot of parameter marginal distributions for a 3 planet HMCMC of the combined Lick, HET, and HJS telescope data set. The residual offset velocity parameters are relative to the Lick dewar 24. They are designated $V_j$, where $j = 6,8,39,18$ correspond to the other Lick dewars and subscripts HET and HJS refer to the Hobby-Eberly Telescope and Harlam J. Smith telescopes \citep{Wittenmyer2009}.}
\label{fig:3planLHETHJSMarg}
\end{figure*}

Figure~\ref{fig:3planLHetEccP} shows a plot of eccentricity versus period for our 3 planet HMCMC fit to the combined data set. The three starting periods used for the HMCMC run were 10, 1078, \& 6000 days. The residual velocity offset parameters for the HET and HJS telescopes were $1.5_{-1.1}^{+1.0}$ and $-0.2\pm 1$ m s$^{-1}$, respectively. 
It is clear from the figure that the same three periods appear as before but with the extra data the results now favor low eccentricity orbits for all three periods. This is a particularly pleasing result as low eccentricity orbits are more likely to exhibit long term stability than high eccentricity orbits.
The preference for low eccentrcity orbits is more apparent in the marginal distributions shown in Figure~\ref{fig:3planLHETHJSMarg}. 

Our final orbital parameters are summarized in Table~\ref{tab:parerrorsM3Combined} together with the residual offset velocities and the extra noise term $s$. Again, the parameter value listed is the median of the marginal probability distribution for the parameter in question and the error bars identify the boundaries of the 68.3\% credible region. The value immediately below in parenthesis is the MAP value, the value at the maximum of the joint posterior probability distribution.
\begin{table}
 \centering
 \begin{minipage}{85mm}
  \caption{Final 3 planet model parameter estimates from the HMCMC fit of the combined Lick, HET, and HJS telescope data set.}
  \label{tab:parerrorsM3Combined}
  \begin{tabular}{@{}lllll@{}}
  \hline
   Parameter  & planet 1 & planet 2 & planet 3 \\
\hline
$P$  (d) & $1078_{-2}^{+2}$ & $2391_{-87}^{+100}$& $14002_{-5095}^{+4018}$  \\
& (1078)& (2430)& (47831)\\
& & &mode$=11251$  \\
& & &  \\
$K$ (m s$^{-1}$) & $48.4_{-0.9}^{+0.8}$ & $8.0_{-1.0}^{+1.0}$ & $13.8_{-2.9}^{+2.2}$  \\
& (48.2) & (8.3) & (13.5) \\
& & &  \\
$e$ & $0.032_{-.014}^{+.014}$ & $0.098_{-.0.096}^{+.0.047}$  & $0.16_{-.16}^{+.09}$   \\
& (0.038) & (0.020) & (0.67) \\
& & &  \\
$\omega$  (deg) & $334_{-23}^{+23}$ & $295_{-160}^{+114}$ &  $110_{-160}^{+132}$  \\
& (324) & (356) & (110) \\
& & &  \\
$a$  (au) & $2.100_{-.02}^{+.02}$ & $3.6_{-.1}^{+.1}$ &  $11.6_{-2.9}^{+2.1}$  \\
& (2.10) & (3.6) &  (26.3)  \\
& & & & \\
$M \sin i$  ($M_J$) & $2.53_{-.06}^{+.07}$ & $0.540_{-.073}^{+.066}$ &  $1.64_{-0.48}^{+.29}$ \\
& (2.53) & (0.567) &  (1.86) \\
& & &  \\
Periastron & $11917_{-76}^{+63}$ & $12441_{-825}^{+628}$ &  $11736_{-5051}^{+6783}$  \\
\ passage &  (11888) & (12778)&  (11736) \\
\ (JD - 2,440,000) & & &  \\
& & & & \\
$V_{6}$  (m s$^{-1}$) & $1.1_{-2.9}^{+2.8}$ & $V_{8}$  (m s$^{-1}$) & $-0.6_{-2.6}^{+2.6}$ \\
& (4.0) &  & (1.0) \\
& & &  \\
$V_{39}$  (m s$^{-1}$) & $-5.0_{-2.7}^{+2.8}$ & $V_{18}$  (m s$^{-1}$) &  $-5.1_{-1.6}^{+1.7}$  \\
& (-0.5) &  & (-4.6) \\
& & &  \\
$V_{HET}$  (m s$^{-1}$) & $1.5_{-1.1}^{+1.0}$ & $V_{HJS}$  (m s$^{-1}$) &  $-0.2_{-1.0}^{+1.0}$  \\
& (1.3) &  & (0.1) \\
& & &  \\
$s$  (m s$^{-1}$) & $5.7_{-0.3}^{+0.3}$ & &    \\
& (5.3) &  &  \\
& & &  \\
\hline
\end{tabular}
\end{minipage}
\end{table}

\begin{figure}
\includegraphics[width=80mm]{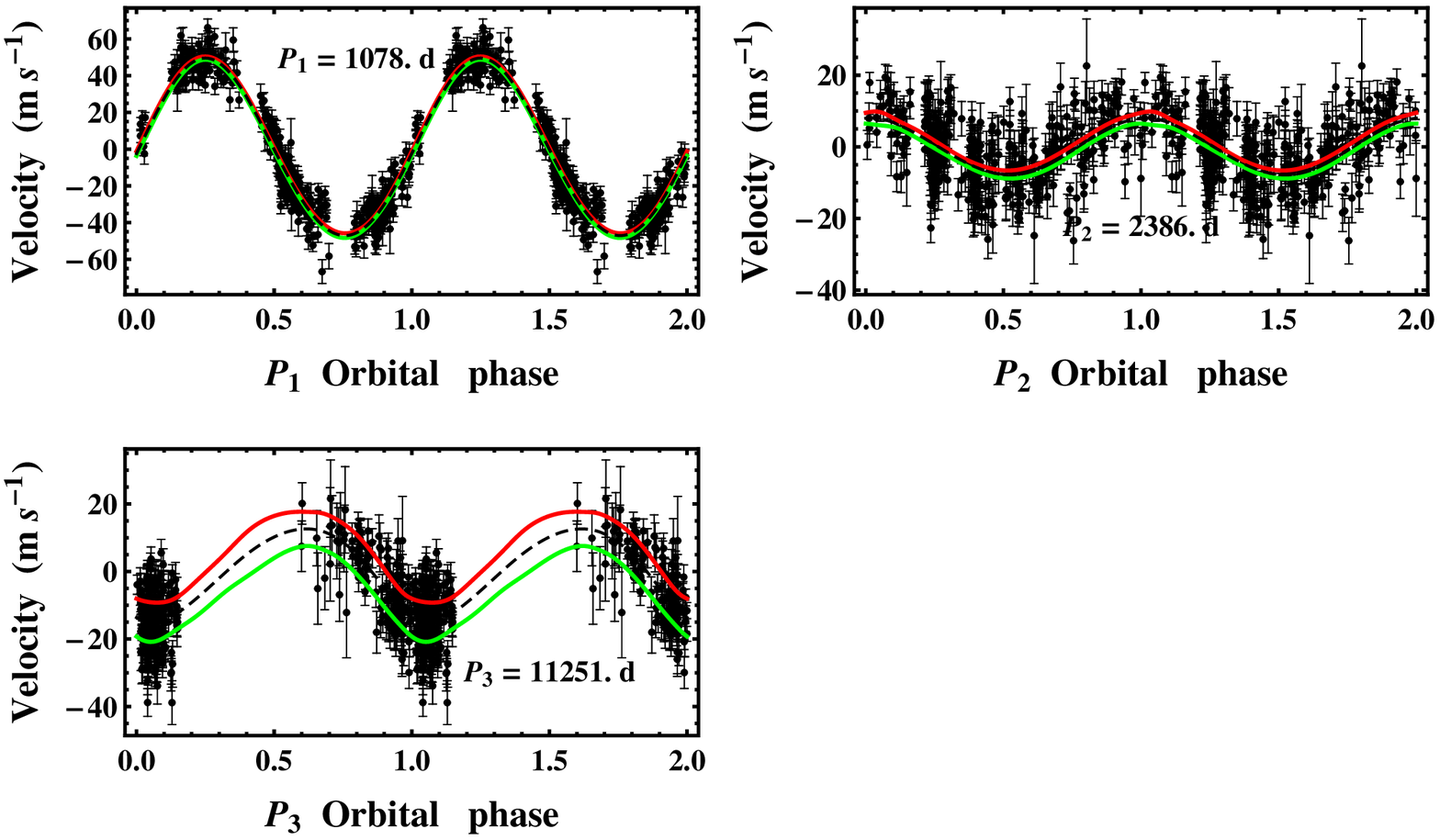}
\caption{The top left panel shows the data and model fit versus 1078 day orbital phase after removing the effects of the two other orbital periods. The red and green curves are the mean HMCMC model fit $+ 1$ standard deviation and mean model fit $- 1$ standard deviation, respectively. The dashed curve is the mean HMCMC fit. The other two panels correspond to phase plot for the other two periods.}
\label{fig:3planPhasePltsCombined}
\end{figure}

The final period phase plots are shown in Figure~\ref{fig:3planPhasePltsCombined}. The top left panel shows the data and model fit versus 1078 day orbital phase after removing the effects of the two other orbital periods. The red and green curves are the mean HMCMC model fit $+ 1$ standard deviation and mean model fit $- 1$ standard deviation, respectively. The dashed curve is the mean HMCMC fit. The other two panels correspond to phase plot for the other two periods. In each panel the quoted period is the mode of the marginal distribution.
The $P_2$ and $P_3$ phase coverage for the combined HET and HJS data (not shown) is not sufficient to warrant a fully independent search for these two periods.

HMCMC fits of a 4 planet model to the combined Lick, HET and HJS data set failed to detect a well defined fourth period casting doubt on the validity of the $370.8_{-2.0}^{+2.4}$ day period detected in the Lick only data. Even though this period was well defined in the Lick only data, the Bayesian false alarm probability of $\approx 0.5$ is much too high to warrant any claim of significance. The period is also suspiciously close to one year and might be an artefact of the data reduction.

\subsection{Eccentricity bias}
\label{sec:eccbias}

In Section~\ref{sec:eccBias} we showed that HMCMC periodogram peaks exhibit a well defined statistical bias towards high eccentricity in the absence of a real periodic signal. To mimic a circular velocity orbit the noise points need to be correlated over a larger fraction of the orbit than they do to mimic a highly eccentric orbit. For this reason it is more likely that noise will give rise to spurious highly eccentric orbits than low eccentricity orbits. Is there a similar or stronger bias when there is a real periodic signal? Based on the above explanation of the bias we would expect noise to conspire to increase the eccentricity of detected periodogram peaks associated with the real periodic signals. Our expectation is that the importance of this bias will be dependent on the strength of the signal and possibly on the number of observed periods~\footnote{This will be the subject of a future investigation.}. For very strong signals like the 1078 day period we would expect the bias to be very small. For very weak signals the bias might well be approximated by the no real periodic signal eccentricity bias which we quantified earlier. As we have seen, in the case of the 47 UMa $\sim 2300$ day period, the Lick data alone favors an eccentricity of $\approx 0.3$, even when we include the eccentricity bias filter. When we added more data the eccentricity was noticeably reduced. What if we simulated a Lick only data set for a 3 planet model based on the MAP 3 planet model parameters for the combined Lick, HET, and HJS analysis. Would the HMCMC analysis of the simulated data favor higher eccentricities, possibly indicating that there is some additional eccentricity bias operating. To test for this we carried out this simulation but modified the MAP parameter values so all three eccentricities were identically zero and $P_3 = 10000$ days. Also, no residual offsets were included for this test so the analysis corresponds to Case A. 
\begin{figure}
\includegraphics[width=80mm]{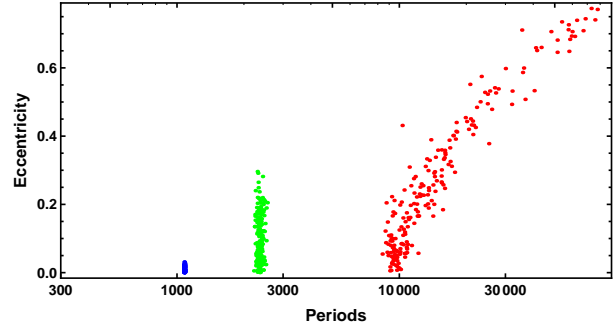}
\caption{A plot of eccentricity versus period for the 3 planet HMCMC fit of the 3 planet simulation.}
\label{fig:3planEccPSimA}
\end{figure} 

Figure~\ref{fig:3planEccPSimA} shows a plot of eccentricity versus period for the simulation. The starting period values for the HMCMC were 5, 20, and 100 days, a long way from the expected values. Again, all three simulated periods were detected and the preferred eccentricities are all close to zero but with significant tails extending to higher eccentricity. Based on this test there does not appear to be any clear additional eccentricity bias operating. The fact that the real Lick data alone favor (in Case A and Case B) somewhat larger eccentricities for $P_2$ and $P_3$ suggests there may be something else present in the real data, possible some low level systematic effect or other real signals. In this regard, the eccentricity of the longer period was considerably higher in the 2 planet models than when allowance was made for an additional period in the 3 planet models.

\section{Conclusions}

In this paper, we have demonstrated that a Bayesian adaptive hybrid MCMC (HMCMC) analysis of a challenging data set has helped clarify the number of planets present in 47UMa. HMCMC integrates the advantages of
parallel tempering, simulated annealing and the genetic algorithm.
Each of these techniques was designed to facilitate the
detection of a global minimum in $\chi^2$. Combining all three
in an adaptive hybrid MCMC greatly increase the probability
of realizing this goal. The adaptive Bayesian hybrid MCMC is very general and
can be applied to many different nonlinear modeling problems. It has been implemented in gridMathematica on an 8 core PC. The increase in a speed for the parallel implementation
is a factor 6.6. When applied to the Kepler problem it corresponds
to a multi-planet Kepler periodogram which is ideally suited for detecting signals that are consistent with Kepler's laws. However, it is more than a periodogram because it also provides full marginal posterior distributions for all the orbital parameters that can be extracted from radial velocity data. The execution time for a 1 planet blind fit (7 parameters) is $10^6$ iterations per hr. The program scales linearly with the number of parameters being fit. 

The 47UMa data has been analyzed using 1, 2, 3, 4, and 5 planet models. On the basis of the model selection results we can conclude there is strong evidence for three planets based on a Bayesian false alarm probability of $5.0 \times 10^{-6}$, however, the longest period orbital parameters are still not well defined. The measured periods (based on the combined data set) are $1078\pm 2$, $2391_{-87}^{+100}$, and $14002_{-5095}^{+4018}$d, and the corresponding eccentricities are $0.032\pm 0.014$, $0.098_{-.096}^{+.047}$, and $0.16_{-.16}^{+.09}$. The results favor low eccentricity orbits for all three. Note: the longer time base of the full Lick data set favors a value for $P_3$ at the lower end of the 68\% credible region of $\sim 10,000$ days. Assuming the three signals (each one consistent with a Keplerian orbit) are caused by planets, the corresponding limits on planetary mass ($M \sin i$) and semi-major axis are 
($2.53_{-.06}^{+.07} M_J$, $2.10\pm 0.02\rm{au}$),\\ ($0.54\pm 0.07 M_J$, $3.6\pm 0.1\rm{au}$), and  ($1.6_{-0.5}^{+0.3} M_J$, $11.6_{-2.9}^{+2.1}\rm{au}$), respectively. Based on our three planet model results, the remaining unaccounted for noise (stellar jitter) is $5.7$m s$^{-1}$.

A four planet model fit to the Lick data yielded a well defined fourth period of $370.8_{-2.0}^{+2.4}$ days and eccentricity of $0.57_{-0.15}^{+0.22}$, but the combined data set did not yield a well defined fourth period. Even though this period was well defined in the Lick only data, the Bayesian false alarm probability of $\approx 0.5$ is much too high to warrant any claim of significance. The period is also suspiciously close to one year and might be an artefact of the data reduction.
 
The velocities of model fit residuals were randomized in multiple trials and processed using a one planet version of the HMCMC Kepler periodogram. In this situation periodogram probability peaks are purely the result of the effective noise. The orbits corresponding to these noise induced periodogram peaks exhibit a well defined statistical bias towards high eccentricity.
We have characterized this eccentricity bias and designed a correction filter that can be used as an alternate prior for eccentricity to enhance the detection of planetary orbits of low or moderate eccentricity. On the basis of our understanding of the mechanism underlying the eccentricity bias, we expect the eccentricity prior filter to be generally applicable to searches for low amplitude orbital signals in other precision radial velocity data sets.
 
\section*{Acknowledgments}

One of us (Gregory) would like to thank Wolfram Research for providing a complementary license to run gridMathematica that was used in this research. D.A.F. acknowledges research support from NASA grant NNX08AF42G. The authors thank Matt Giguere for calculating dewar offsets and
the many student observers who have contributed to the 47 UMa data
set over the years. They particularly thank Geoff Marcy and R. Paul Butler
who began the Lick planet survey in 1989.

\bsp

\label{lastpage}


\section{Bibliography}

\begin{thebibliography}{99}

\bibitem[\protect\citeauthoryear{Bretthorst}{1988}]{Brett1988} Bretthorst, G. L., 1988, Bayesian Spectrum Analysis and Parameter
Estimation, New York: Springer-Verlag




\bibitem[\protect\citeauthoryear{Butler \& Marcy}{1996}]{ButlerMarcy1996} Butler, R. P. \& Marcy, G. W. 1996, ApJ, 464, L153


\bibitem[\protect\citeauthoryear{Cumming}{2004}]{Cumming2004} Cumming, A., 2004, MNRAS, 354, 1165

\bibitem[\protect\citeauthoryear{Cumming \& Dragomir}{2010}]{CummingDragomir2010} Cumming, A., Dragomir, D., 2010, MNRAS, 401, 1029

\bibitem[\protect\citeauthoryear{Clyde et al.}{2006}]{Clyde2006} Clyde, M. A., Berger, J. O., Bullard, F., Ford, E. B., Jeffreys, W. H., Luo, R., Paulo, R., Loredo, T., 2006, in `Statistical Challenges in Modern Astronomy IV,'  G. J. Babu and E. D. Feigelson (eds.), ASP Conf. Ser., 371, 224 
 
 

\bibitem[\protect\citeauthoryear{Fischer et al.}{2002}]{Fischer2002} Fischer,D. A., Marcy, G. W., Butler, R. P., Laughlin, G. L., and Vogt, S. S., 2002, ApJ, 564, 1028

\bibitem[\protect\citeauthoryear{Ford}{2005}]{Ford2005} Ford, E. B., 2005, AJ, 129, 1706

\bibitem[\protect\citeauthoryear{Ford}{2006}]{Ford2006} Ford, E. B., 2006, ApJ, 620, 481

\bibitem[\protect\citeauthoryear{Ford \& Gregory}{2006}]{FordGregory2006} Ford, E. B., \& Gregory, P. C., 2006, in `Statistical Challenges in Modern Astronomy IV,'  G. J. Babu and E. D. Feigelson (eds.), ASP Conf. Ser., 371, 189   
      
 
\bibitem[\protect\citeauthoryear{Gelman-Rubin}{1992}]{Gel} Gelman, A., \& Rubin, D. B., 1992, Statistical Science 7, 457

\bibitem[\protect\citeauthoryear{Gregory \& Loredo}{1992}]{GregoryLoredo1992} Gregory, P. C., and Loredo, T. J., 1992, ApJ, 398, 146

\bibitem[\protect\citeauthoryear{Gregory}{2005a}]{Gregorybook} Gregory, P. C., 2005a, `Bayesian Logical Data Analysis for the Physical Sciences: A Comparative approach with {\it Mathematica} Support', Cambridge University Press

\bibitem[\protect\citeauthoryear{Gregory}{2005b}]{Gregory2005b} Gregory, P. C., 2005b, ApJ, 631, 1198

\bibitem[\protect\citeauthoryear{Gregory}{2005c}]{Gregory2005MaxEnt} Gregory, P. C.,2005c, in `Bayesian Inference and Maximum Entropy Methods in Science and Engineering', San Jose, eds. A. E. Abbas, R. D. Morris, J. P.Castle, AIP Conference Proceedings, 803, 139

\bibitem[\protect\citeauthoryear{Gregory}{2007a}]{Gregory2007a} Gregory, P. C., 2007a, MNRAS, 374, 1321

\bibitem[\protect\citeauthoryear{Gregory}{2007b}]{Gregory2007b} Gregory, P. C., 2007b, in `Bayesian Inference and Maximum Entropy Methods in Science and Engineering: 27th International Workshop', Saratoga Springs, eds. K. H. Knuth, A. Caticha, J. L. Center, A,\. Giffin, C. C. Rodríguez, AIP Conference Proceedings, 954, 307

\bibitem[\protect\citeauthoryear{Gregory}{2007c}]{Gregory2007c} Gregory, P. C., 2007c, MNRAS, 381, 1607

\bibitem[\protect\citeauthoryear{Gregory}{2009}]{Gregory2009} Gregory, P. C., 2009, JSM Proceedings, Denver, American Statistical Association, arXiv:0902.2014v1 [astro-ph.EP]

\bibitem[\protect\citeauthoryear{Jaynes}{1957}]{Jaynes1957} Jaynes, E. T., 1957, Stanford University Microwave Laboratory Report 421, Reprinted in `Maximum Entropy and Bayesian Methods in Science and Engineering', G. J. Erickson and C. R. Smith, eds, (1988) Dordrecht: Kluwer Academic Press, p.1

\bibitem[\protect\citeauthoryear{Jaynes}{1987}]{Jaynes1987} Jaynes, E.T. (1987), `Bayesian Spectrum \& Chirp Analysis,' in {\it Maximum Entropy and Bayesian Spectral Analysis and Estimation Problems}, ed. C.R. Smith and G.L. Erickson, D. Reidel, Dordrecht, p. 1


\bibitem[\protect\citeauthoryear{Loredo}{2004}]{Loredo2004} Loredo, T., 2004, in `Bayesian Inference And Maximum Entropy Methods in Science and Engineering: 23rd International Workshop', G.J. Erickson \& Y. Zhai, eds, AIP Conf. Proc. 707, 330 (astro-ph/0409386)

\bibitem[\protect\citeauthoryear{Loredo \& Chernoff}{2003}]{LoredoChernoff2003} Loredo, T. L. and Chernoff, D., 2003, in `Statistical Challenges in Modern Astronomy III', E. D. Feigelson and G. J. Babu (eds) , Springer, New York, p. 57

\bibitem[\protect\citeauthoryear{Naef et al.}{2004}]{Naef2004} Naef, D.,Mayor,M., Beuzit, J. L., Perrier, C., Queloz, D., Sivan, J. P., \& Udry, S.
2004, A\&A, 414, 351

  
\bibitem[\protect\citeauthoryear{Roberts et al.}{1997}]{Roberts1997} Roberts, G. O., Gelman, A. and Gilks, W. R., 1997, Annals of Applied Probability, 7, 110

\bibitem[\protect\citeauthoryear{Saar \& Donahue}{1997}]{Saar1997} Saar, S. H., \& Donahue, R. A., 1997, ApJ, 485, 319

\bibitem[\protect\citeauthoryear{Saar et al.}{1998}]{Saar1998} Saar, S. H., Butler, R. P., \& Marcy, G. W. 1998, ApJ, 498, L153




\bibitem[\protect\citeauthoryear{Takeda et al.}{2007}]{Takeda2007} Takeda, G., Ford, E. B., Sills, A., Rasio, F. A., Fischer, D. A., \& Valenti, J. A. 2007, ApJS,
168, 297


\bibitem[\protect\citeauthoryear{Wittenmyer, Endl \& Cochran}{2007}]{Wittenmyer2007}{Wittenmyer, R., Endl, M. \& Cochran, W., 2007, ApJ, 654, 625}

\bibitem[\protect\citeauthoryear{Wittenmyer et al.}{2009}]{Wittenmyer2009}{Wittenmyer, R. A., Endl, M., Cochran, W. D., Levison, H. F., Henry, G. W., 2009, ApJS, 182, 97}

\bibitem[\protect\citeauthoryear{Wright}{2005}]{Wright2005} Wright, J. T., 2005, PASP, 117, 657


\bibitem[\protect\citeauthoryear{Udry et al.}{2007}]{Udry2007}{Udry, S., Bonfils, X., Delfosse, X., Forveille, T., Mayor, M., Perrier, C., Bouchy, F., Lovis, C., Pepe, F., Queloz, D., and Bertaux, J.-L., 2007, A\&A, 469, L43}

\end{thebibliography}
\end{document}